\documentclass[acmsmall, nonacm=true]{acmart}

\usepackage{import}

\usepackage{xspace}

\newcommand{\Ttwo}{T_2}

\newcommand{\qwire}{\ensuremath{\mathcal{Q}\textsc{wire}}\xspace}
\newcommand{\sqir}{s\textsc{qir}\xspace}
\newcommand{\liquid}{LIQ$Ui\ket{}$\xspace}

\usepackage{xcolor}

\definecolor{eclipseBlue}{RGB}{42,0.0,255}
\definecolor{eclipseGreen}{RGB}{63,127,95}
\definecolor{eclipsePurple}{RGB}{127,0,85}
\definecolor{eclipseRed}{RGB}{223,26,65}
\definecolor{sublimegreen}{RGB}{112,200,0}
\definecolor{purple}{RGB}{255, 0, 255}
\definecolor{qing}{RGB}{32, 101, 159}

\definecolor{grayonine}{gray}{0.9}
\definecolor{grayoeight}{gray}{0.8}
\definecolor{grayoseven}{gray}{0.7}
\definecolor{grayosix}{gray}{0.6}
\definecolor{grayofive}{gray}{0.5}
\definecolor{grayofour}{gray}{0.4}
\definecolor{grayothree}{gray}{0.3}
\definecolor{grayotwo}{gray}{0.2}
\definecolor{grayoone}{gray}{0.1}

\definecolor{opcodecolor}{RGB}{0,0,0}
\definecolor{feedlinegray}{gray}{0.8}
\definecolor{codegray}{gray}{0.9}

\usepackage{amssymb}
\usepackage{amsmath}

\usepackage{braket}

\usepackage{url}

\usepackage{natbib}

\usepackage{enumitem}

\usepackage{graphicx}
\usepackage{pythonhighlight}
\usepackage{listings}

\usepackage{tikz}
\usetikzlibrary{decorations.pathreplacing,decorations.pathmorphing}

\usepackage{siunitx}

\usepackage{parcolumns}

\newcommand{\code}[1]{{\small \texttt{#1}}}

\newcommand{\kw}[1]{{\textit{#1}}}        % keyword

\newcommand{\lines}[2]{lines~#1\ --\ #2\xspace}

\newcommand{\ipeonecontent}{Quantum kernel preparation}
\newcommand{\ipethreecontent}{Post-processing}
\newcommand{\ipetwocontent}{Quantum execution}
\newcommand{\ipefourcontent}{Search}
\newcommand{\ipefivecontent}{Repetition \& optimization}

\newcommand{\ipeone}{\kw{\ipeonecontent}\xspace}
\newcommand{\ipetwo}{\kw{\ipetwocontent}\xspace}
\newcommand{\ipethree}{\kw{\ipethreecontent}\xspace}
\newcommand{\ipefour}{\kw{\ipefourcontent}\xspace}
\newcommand{\ipefive}{\kw{\ipefivecontent}\xspace}

\usepackage{float}
\newfloat{lstfloat}{htbp}{lop}
\floatname{lstfloat}{Listing}
\usepackage{listings}

\lstdefinelanguage{eQASM}
{
  sensitive=false, % keywords are not case-sensitive
  alsoletter={.1234567890},
  keywords={},
  otherkeywords={ % operators
  |
  },
  morekeywords=[2]{always, never, eq, ne, eqz, nez, lt, ltz, le, gt, ge, gez, ltu, leu, gtu, geu, carry, notcarry},
  morekeywords=[3]{qwait, qwaitr, SMIS, SMIT},
  morekeywords=[4]{bs},
  morekeywords=[5]{MeasZ, X, Y, CNOT, X180, Xm180, X90, Xm90, Y180, Ym180, Ym90, Y90, CZ, T, H, S, Z, QNOP},
  morekeywords=[6]{nop, stop, ldi, ldui, add, addi, sw, lw, fcvt.s.w, sub, addc, subc, and, or, xor, not, fmr, br, cmp, fbr,  st, ld},
  morekeywords=[7]{.register, .def_sym},
  morekeywords=[8]{nand, nor, xnor, bra, brn, beq, bne, blt, ble, bgt, bge, bltu, bleu, bgtu, bgeu, goto, mov, shl1, mult2},
  morecomment=[l]{\#}, % l is for line comment
}

\lstdefinestyle{eQASMStyle}{
  language=eQASM,
  linewidth=0.94\columnwidth,
  xleftmargin=0.06\columnwidth,
  basicstyle=\linespread{1.2}\scriptsize\ttfamily, % Global Code Style
  captionpos=t, % Position of the Caption (t for top, b for bottom)
  extendedchars=true, % Allows 256 instead of 128 ASCII characters
  tabsize=2, % number of spaces indented when discovering a tab 
  columns=fixed, % make all characters equal width
  keepspaces=true, % does not ignore spaces to fit width, convert tabs to spaces
  breaklines=true, % wrap lines if they don't fit
  frame=trbl, % draw a frame at the top, right, left and bottom of the listing, other options: single, shadowbox
  rulecolor=\color{black!40},
  numbers=left, % show line numbers at the left
  commentstyle=\color{eclipseGreen}, % style of comments
  keywordstyle=\color{eclipseGreen}, % style of keywords
  keywordstyle=[2]*\bfseries\color{black}, % style of keywords
  keywordstyle=[3]*\color{sublimegreen}, % * means capitalize 
  keywordstyle=[4]\color{grayosix},
  keywordstyle=[5]*\bfseries\color{orange}, % * means capitalize 
  keywordstyle=[6]*\color{eclipseBlue}, % * means capitalize 
  keywordstyle=[7]\color{eclipsePurple}, % * means capitalize 
  keywordstyle=[8]*\color{eclipseRed}
}

\lstnewenvironment{eQASM}%
{\lstset{style=eQASMStyle}}
{}
\RequirePackage{listings}
\RequirePackage{xcolor}

\colorlet{stringcolour}{red!60!black}
\colorlet{keywordcolour}{magenta!90!black}
\colorlet{exceptioncolour}{yellow!50!red}
\colorlet{commandcolour}{blue!60!black}
\colorlet{numpycolour}{blue!60!green}
\colorlet{literatecolour}{magenta!90!black}
\colorlet{promptcolour}{green!50!black}
\colorlet{specmethodcolour}{violet}

\lstdefinelanguage{Quingo}
{
  sensitive=true, % keywords are case-sensitive
  alsoletter={.1234567890},
  keywords={},
  otherkeywords={ % operators
  |
  },
  morekeywords=[2]{import, operation, opaque, package},
  morekeywords=[3]{using},
  morekeywords=[4]{PI, ps, ns, us, ms, sec},
  morekeywords=[5]{I, MeasZ, X, Y, CNOT, X180, Xm180, X90, Xm90, Y180, Ym180, Ym90, Y90, CZ, T, Tdagger, Tdag, H, S, Z, QNOP, init, measure, RX, RZ, RY},
  morekeywords=[6]{qubit, double, int, bool, unit},
  morekeywords=[7]{time, timer},
  morekeywords=[8]{while, for, if, else, return, continue, break, switch, control, invert},
  morekeywords=[9]{false, true},
  morecomment=[l]{//}, % l is for line comment
}

\lstdefinestyle{QuingoStyle}{
  language=Quingo,
  linewidth=0.94\columnwidth,
  xleftmargin=0.06\columnwidth,
  basicstyle=\linespread{1.0}\scriptsize\ttfamily, % Global Code Style
  captionpos=t, % Position of the Caption (t for top, b for bottom)
  extendedchars=true, % Allows 256 instead of 128 ASCII characters
  tabsize=2, % number of spaces indented when discovering a tab
  columns=fixed, % make all characters equal width
  keepspaces=true, % does not ignore spaces to fit width, convert tabs to spaces
  breaklines=true, % wrap lines if they don't fit
  frame=trbl, % draw a frame at the top, right, left and bottom of the listing, other options: single, shadowbox
  rulecolor=\color{black!40},
  numbers=left, % show line numbers at the left
  commentstyle=\color{eclipseGreen}, % style of comments
  keywordstyle=\color{eclipseGreen}, % style of keywords
  keywordstyle=[2]\bfseries\color{keywordcolour}, % style of keywords
  keywordstyle=[3]\color{sublimegreen}, % * means capitalize
  keywordstyle=[4]\color{numpycolour},
  keywordstyle=[5]\bfseries\color{orange}, % * means capitalize
  keywordstyle=[6]\color{eclipseBlue}, % * means capitalize
  keywordstyle=[7]\color{eclipsePurple}, % * means capitalize
  keywordstyle=[8]\color{eclipseRed},
  keywordstyle=[9]\bfseries\color{sublimegreen}
}

\lstnewenvironment{Quingo}[1][]{\lstset{style=QuingoStyle}}{}

\lstdefinestyle{StyleQuingoinline}{
  language=Quingo,
  linewidth=0.95\columnwidth,
  xleftmargin=0.06\columnwidth,
  basicstyle=\linespread{1}\small\ttfamily, % Global Code Style
  captionpos=b, % Position of the Caption (t for top, b for bottom)
  extendedchars=true, % Allows 256 instead of 128 ASCII characters
  tabsize=2, % number of spaces indented when discovering a tab
  columns=fixed, % make all characters equal width
  keepspaces=true, % does not ignore spaces to fit width, convert tabs to spaces
  breaklines=true, % wrap lines if they don't fit
  frame=trbl, % draw a frame at the top, right, left and bottom of the listing, other options: single, shadowbox
  frameround=tttt, % make the frame round at all four corners
  numbers=left, % show line numbers at the left
commentstyle=\color{eclipseGreen}, % style of comments
keywordstyle=\color{eclipseGreen}, % style of keywords
keywordstyle=[2]\bfseries\color{keywordcolour}, % style of keywords
keywordstyle=[3]\color{sublimegreen}, % * means capitalize
keywordstyle=[4]\color{numpycolour},
keywordstyle=[5]\bfseries\color{orange}, % * means capitalize
keywordstyle=[6]\color{eclipseBlue}, % * means capitalize
keywordstyle=[7]\color{eclipsePurple}, % * means capitalize
keywordstyle=[8]\color{eclipseRed},
keywordstyle=[9]\bfseries\color{eclipseRed}
}

\newcommand{\quingo}{\lstinline[style=StyleQuingoinline]}

\begin{document}

\title{Quingo: A Programming Framework for Heterogeneous Quantum-Classical Computing with NISQ Features}
\author{The Quingo Development Team:~~~X. Fu}
\affiliation{%
  \institution{Institute for Quantum Information \& State Key Laboratory of High Performance Computing, College of Computer, National University of Defense Technology}
  \city{Changsha}
  \postcode{410073}
  \country{China}
}
\author{Jintao Yu}
\affiliation{%
  \institution{State Key Laboratory of Mathematical Engineering and Advanced Computing}
  \city{Zhengzhou}
  \postcode{450001}
  \country{China}
}
\author{Xing Su}
\affiliation{%
  \institution{College of Computer, National University of Defense Technology}
  \city{Changsha}
  \postcode{410073}
  \country{China}
}
\author{Hanru Jiang}
\affiliation{%
  \institution{Center for Quantum Computing, Peng Cheng Laboratory}
  \city{Shenzhen}
  \postcode{518055}
  \country{China}
}
\author{Hua Wu}
\affiliation{%
  \institution{Shanghai Key Laboratory of Trustworthy Computing, East China Normal University}
  \city{Shanghai}
  \postcode{200062}
  \country{China}
}
\author{Fucheng Cheng}
\author{Xi Deng}
\author{Jinrong Zhang}
\affiliation{%
  \institution{Center for Quantum Computing, Peng Cheng Laboratory}
  \city{Shenzhen}
  \postcode{518055}
  \country{China}
}
\author{Lei Jin}
\author{Yihang Yang}
\author{Le Xu}
\author{Chunchao Hu}
\affiliation{%
  \institution{School of Information Engineering, Zhengzhou University}
  \city{Zhengzhou}
  \postcode{450001}
  \country{China}
}
\author{Anqi Huang}
\author{Guangyao Huang}
\author{Xiaogang Qiang}
\author{Mingtang Deng}
\author{Ping Xu}
\author{Weixia Xu}
\affiliation{%
  \institution{Institute for Quantum Information \& State Key Laboratory of High Performance Computing, College of Computer, National University of Defense Technology}
  \city{Changsha}
  \postcode{410073}
  \country{China}
}
\author{Wanwei Liu}
\affiliation{%
  \institution{Department of Computing Science, College of Computer, National University of Defense Technology}
  \city{Changsha}
  \postcode{410073}
  \country{China}
}
\author{Yu Zhang}
\affiliation{%
  \institution{School of Computer Science and Technology, University of Science and Technology of China}
  \city{Hefei}
  \postcode{230027}
  \country{China}
}
\author{Yuxin Deng}
\email{yxdeng@sei.ecnu.edu.cn}
\affiliation{%
  \institution{Shanghai Key Laboratory of Trustworthy Computing, East China Normal University}
  \city{Shanghai}
  \postcode{200062}
  \country{China}
}
\author{Junjie Wu}
\email{junjiewu@nudt.edu.cn}
\affiliation{%
  \institution{Institute for Quantum Information \& State Key Laboratory of High Performance Computing, College of Computer, National University of Defense Technology}
  \city{Changsha}
  \postcode{410073}
  \country{China}
}
\author{Yuan Feng}
\email{yuan.feng@uts.edu.au}
\affiliation{%
  \institution{ Centre for Quantum Software and Information, University of Technology Sydney}
  \city{Sydney}
  \postcode{2007}
  \country{Australia}
}

\begin{abstract}

The increasing control complexity of Noisy Intermediate-Scale Quantum (NISQ) systems underlines the necessity of integrating quantum hardware with quantum software. While mapping heterogeneous quantum-classical computing (HQCC) algorithms to NISQ hardware for execution, we observed a few dissatisfactions in quantum programming languages (QPLs), including difficult mapping to hardware, limited expressiveness and counter-intuitive code. Also, noisy qubits require repeatedly-performed quantum experiments, which explicitly operate low-level configurations, such as pulses and timing of operations. This requirement is beyond the scope or capability of most existing QPLs.

We summarize three execution models to depict the quantum-classical interaction of existing QPLs. Based on the refined HQCC model, we propose the Quingo framework to integrate and manage quantum-classical software and hardware to provide the programmability over HQCC applications and map them to NISQ hardware. We propose a six-phase quantum program life-cycle model matching the refined HQCC model, which is implemented by a runtime system. We also propose the Quingo programming language, an external domain-specific language highlighting timer-based timing control and opaque operation definition, which can be used to describe quantum experiments. We believe the Quingo framework could contribute to the clarification of key techniques in the design of future HQCC systems.
\end{abstract}

\renewcommand{\shortauthors}{The Quingo Development Team}

\keywords{Quantum Programming Framework, Quantum Programming Language, Quantum Compilation, NISQ, Timing Control}

\maketitle

\section{Introduction}
\label{sec:intro}

The potential in solving classically intractable problems such as decryption and quantum chemistry simulation has attracted intensive research on quantum computing.
Advancement in quantum information theory, quantum hardware, and quantum control, contributed to the arrival of the Noisy Intermediate-Scale Quantum (NISQ)~\cite{preskill2012quantum, arute2019quantum, zhong2020quantum} era. In the NISQ era, a quantum system can integrate dozens of noisy qubits with limited coherence time and support complex quantum-classical interaction, such as real-time feedback based on the measurement of qubits~\cite{ryan2017hardware, salathe2018low, fu2019eqasm, corcoles2021exploiting}.
With the increased number of qubits and enhanced control capability, the control complexity of executing quantum applications on NISQ systems grows significantly, stressing the importance of high-level quantum programming languages~\cite{heim2020quantum} in describing various quantum applications in the NISQ era including algorithms as well as experiments.
While being connected to NISQ hardware, existing quantum programming frameworks and languages suffer from difficult mapping to nowadays heterogeneous quantum-classical architectures or tedious and error-prone description of quantum applications, and limited capability in describing NISQ experiments that occupies most of the qubit usage time in the NISQ era.

\subsection{Support for Heterogeneous Quantum-Classical Computation}

It has been shown that practical quantum computing relies on the synergy between quantum and classical computing resources, and it is a viable way to take quantum computers as coprocessors of classical computing systems in a heterogeneous system.

Since Knill proposed the Quantum Random Access Machine (QRAM) model [Fig.~\ref{fig:models}(a)] in 1996~\cite{knill1996conventions}, dozens of quantum programming languages have been proposed to describe heterogeneous quantum-classical computation (HQCC).
While mapping quantum algorithms described by these quantum programming languages to NISQ hardware for execution, they are confronted with three kinds of problems.
\begin{itemize}[leftmargin=*]
    \item  %
The QRAM model is a neat model abstracting away implementation details.
Quantum programming languages directly based on the QRAM model, such as QCL~\cite{omer2003structured} and Scaffold~\cite{javadiabhari2012scaffold}, allow arbitrarily complex classical computation inserted during quantum state evolution subject to applied quantum operations. As NISQ qubits have limited coherence time, it may fail when mapping programs described in these quantum programming languages to hardware for execution (see Section~\ref{sec:models} for a detailed discussion).
    \item %
With the limited qubit coherence time in mind, some quantum programming languages, such as Cirq~\cite{cirq}, are designed with strict constraints put on real-time classical computation and quantum-classical interaction.
Although these constraints enable a more reliable mapping of quantum algorithms to real hardware, they are over strong and disable describing features that have been demonstrated by real hardware such as real-time feedback based on the measurement of qubits~\cite{heim2020quantum}.
    \item %
Another set of quantum programming languages can support hardware-implementable (real-time) quantum-classical interaction with reasonable constraints. They are implemented as a Domain-Specific Language (DSL) embedded in a classical language, such as PyQuil~\cite{pyquil} embedded in Python and OpenQL~\cite{khammassi2020openql} in C++.
However, meta-programming techniques need to be used to describe real-time quantum-classical interaction, resulting in counter-intuitive and complicated code, especially when control-flow structures are involved.
Take the PyQuil code as shown in Code~\ref{code:pyquil_loop} as an example. It repeatedly prepares a qubit to the state $\frac{1}{\sqrt{2}}(\ket{0}-\ket{1})$ and measures it until the measurement result is 0 ($\ket{0}$). To support real-time program flow control based on the measurement of a qubit, it requires to use dedicated \code{while\_do} function that takes as parameters a low-level \code{BIT}-type value and a \code{Program}-type object composed of sub-circuits. However, it is preferable to implement the same logic with only high-level programming constructs in a much neater way, such as Code~\ref{code:quingo-loop} (detailed in Section~\ref{sec:user-defined}).
\end{itemize}

Unsatisfactory support for programming HQCC applications targeting execution on NISQ hardware is an issue that should be addressed.

\begin{lstfloat}[bt]
\begin{lstlisting}[
  style=mypython,
  numbers=left,
  linewidth=0.94\columnwidth,
  xleftmargin=0.06\columnwidth,
  basicstyle=\linespread{1}\scriptsize\ttfamily,
  caption=PyQuil code implementing real-time flow control based on the measurement of a qubit.,
  label=code:pyquil_loop,
]
from pyquil import Program
from pyquil.gates import *

main_prog = Program()  # Initialize the Program
reg_flag  = main_prog.declare('reg_flag', 'BIT')   # declare a 1 bit memory

main_prog += MOVE(reg_flag, 1)           # initial reg_flag to 1

inner_loop = Program()                    # Define the body of the loop
inner_loop += Program(X(0), H(0))
inner_loop += MEASURE(0, reg_flag)

main_prog.while_do(reg_flag, inner_loop) # loop until reg_flag is 0

print(main_prog)
\end{lstlisting}

\begin{lstlisting}[
  style=QuingoStyle,
  numbers=left,
  linewidth=0.94\columnwidth,
  xleftmargin=0.06\columnwidth,
  basicstyle=\linespread{1}\scriptsize\ttfamily,
  caption=A neater way to implement real-time flow control based on the measurement of a qubit with only high-level programming constructs.,
  label=code:quingo-loop,
]
import config.json

operation loop(): unit {
    bool flag = true;
    using (q: qubit) {
        while (flag) {
            RX(q, PI);
            H(q);
            flag = measure(q);
        }
    }
}
\end{lstlisting}
\end{lstfloat}

\subsection{Support for Quantum Experiments}
The noisy nature of qubits makes it essential to repeatedly perform quantum experiments calibrating qubits and tuning quantum operations.
Now and in the foreseeable future, quantum experiments would occupy most of the qubit usage time in the NISQ era.
While performing quantum experiments, experimentalists have to explicitly operate some low-level configurations, such as applying pulses without well-defined quantum semantics and tuning the timing of pulses. For example, a set of pulses with the same envelope but different amplitudes are used in the Rabi experiment to calibrate $x$- and/or $y$-rotations with particular rotation angles~\cite{rabi1937space, reed2013entanglement}. In the experiment measuring the qubit dephasing time ($\Ttwo$), the intervals between the $X_{\pi/2}$ operations must be changed explicitly.

Most of the existing quantum programming languages aim to provide high-level features to support efficient description, optimization, and verification of quantum algorithms.
They intentionally abstract away hardware-dependent constraints.
For example, Q\# targets large-scale applications on future quantum hardware and is hardware-agnostic~\cite{svore2018q, heim2020quantum}.
As a result, operating low-level hardware configuration required by quantum experiments is beyond the scope or capability of most existing quantum programming languages.
Consequently, dedicated experiment environments in a classical programming language such as QCoDeS~\cite{qcodes} or PycQED~\cite{pycqed} are used by experimentalists (referred to as \textit{experiment toolchain}).
After qubits and quantum operations have been calibrated in this environment, the right hardware configuration and pulse parameters can be determined. They are later used by a hardware-aware conversion layer to convert compiled quantum applications to the control electronics input~\cite{chong2017programming} (referred to as \textit{application toolchain}). In this way, some quantum applications described in high-level quantum programming languages can be executed on physical qubits.

Though workable, there are some drawbacks to using two toolchains to interact with qubits.
First, two toolchains can result in an undesired capability mismatch between them.
On the one hand, features well supported by the experiment toolchain may not be expressed in the application toolchain.
For example, feedback based on measurement results can be supported by many experiment setups but cannot be expressed by some high-level quantum programming languages, such as Cirq~\cite{cirq}.
On the other hand, some features can be easily expressed in the application toolchain, but are difficult to be converted to the format accepted by the experiment toolchain for hardware execution, \textit{even if the hardware is capable of supporting this feature}.
For example, although Scaffold~\cite{javadiabhari2012scaffold} can support program flow control based on the measurement result, a feature also supported by some hardware backends~\cite{ryan2017hardware, fu2019eqasm, salathe2018low, corcoles2021exploiting}, there is a difficulty for its compiler ScaffCC~\cite{javadiabhari2015scaffcc} in generating the accepted code with real-time control flow, as mentioned in~\cite{heim2020quantum, scaffccfeedback}.
Second, as long as the hardware-aware conversion between the compiler and hardware is still required, quantum compilers can hardly perform thorough platform-specific optimization on all possible degrees of freedom as some low-level details are hidden from the compiler.

Required is a neat quantum programming framework and language with proper assumptions on the quantum-classical interaction, which describe HQCC algorithms and quantum experiments in a format that can be easily mapped to NISQ hardware for execution. Hopefully, this framework can satisfy the requirements of both the algorithm toolchain and experiment toolchain and bridges the gap in the between.

\subsection{Contribution}

In order to address this issue, we propose the Quingo
(pronunciation: \raisebox{-0.65ex}{\includegraphics[scale=1]{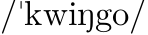}}) framework, a proposal to integrate and manage quantum-classical software and hardware to provide the programmability over HQCC applications and quantum experiments that can be mapped to NISQ hardware.

The main contributions of this paper are as follows:
\begin{itemize}
\item
We summarize three execution models (Section~\ref{sec:models}), viz., the QRAM model, the restricted HQCC model, and the refined HQCC model, to depict the quantum-classical interaction observed in existing quantum programming languages. Compared to the other two models, quantum languages based on the refined HQCC model can be expressive and enable a neat description of quantum-classical interaction more suitable to be mapped to the NISQ hardware for execution.

\item Aiming to serve quantum computing in the NISQ era based on the refined HQCC model, we propose the Quingo framework at a system level to integrate and manage quantum-classical software and hardware to provide programmability over HQCC applications and experiments and map them to NISQ hardware.

\item  We propose a six-phase quantum program life-cycle model, which can guide the development, compilation, optimization, and execution of HQCC applications. This model defines how the Quingo framework integrates and manages quantum and classical computing resources.

\item The common services of the Quingo framework are provided by the Quingo runtime system according to the six-phase quantum program life-cycle model. We have implemented an open-source prototype of the runtime system in Python, which can orchestrate both quantum and classical software and hardware. It allows components of the framework to focus on their genuine tasks, accordingly achieving a modular programming framework. In particular, we define a protocol for the interaction between the classical host\footnote{
It is worth noting that the term `host' is used with two different meanings throughout this paper.
First, when we talk about embedded domain-specific languages, the \textit{host} language refers to the general-purpose language that implements the DSL.
For example, Python is the host language of Qiskit.
Second, when we talk about heterogeneous computing, \textit{host} and \textit{kernel} refer to the language, program, compiler, or hardware corresponding to the general-purpose part and the coprocessor part, respectively.
This pair of terms are borrowed from the OpenCL framework.
The readers should pay attention to which meaning is used according to the context.} language and the quantum kernel language, which enables connecting an external DSL for HQCC, like Quingo or Q\#, to any general-purpose classical programming language, such as Python or C++.

\item To support quantum experiments, we propose the Quingo language, which is an external DSL\footnote{An external DSL is completely designed and implemented from the scratch, not relying on any pre-existent language~\cite{Oliveira2009domain}.} for quantum computing. The Quingo language highlights (1) a flexible, timer-based timing control scheme at the language level with well-defined semantics, and (2) a mechanism for primitive operation definition that enables a flexible binding between operations at the language level and concrete semantics (unitaries or pulses) on the target platform. We have implemented a prototype compiler that can handle quantum-classical interaction and translate the Quingo program into eQASM instructions.

\end{itemize}

This paper is organized as follows. After illustrating the structure of HQCC algorithms with an example, Section~\ref{sec:background} presents and compares the three execution models summarized for HQCC.
Section~\ref{sec:overview} gives an overview of the Quingo framework based on the refined HQCC model, whose key techniques are presented in Section~\ref{sec:hqcc}.
Section~\ref{sec:language} introduces the Quingo language  for describing quantum kernels of both quantum algorithms and experiments.
Quingo-related implementation as well as an example is shown in Section~\ref{sec:example}.
After discussing some design choices and the rationales of the Quingo framework in Section~\ref{sec:rr}, we conclude in Section~\ref{sec:conclusion}.
\section{Background}
\label{sec:background}

\subsection{HQCC Algorithms}
\label{sec:vqe_example}
Able to dramatically reduce the requirement on the number of qubits and the qubit coherence time, algorithms that utilize both quantum and classical computing are highly promising in the near term with wide applications in quantum simulation and optimization.
Quantum phase estimation (QPE) is a key component for a wide range of applications, such as quantum simulation~\cite{peruzzo2014variational, omalley2016scalable} and Shor's factoring~\cite{shor1994algorithms}, and a flavor that utilizes both quantum-classical computation, namely iterative phase estimation (IPE), has been proposed recently.
Compared with other flavors of phase estimation such as Kitaev QPE~\cite{kitaev1995quantum}, the measurement results help IPE to avoid intensive classical computing.
We take IPE as an example to illustrate some properties of HQCC algorithms and show their requirement on both the language and the programming framework.
This section briefly introduces the principle and procedure of the IPE algorithm, and interested readers are referred to \cite{dobvsivcek2007arbitrary, svore2013faster, corcoles2021exploiting} for more details.

\begin{figure}[bt]
    \centering
    \includegraphics[width=\columnwidth]{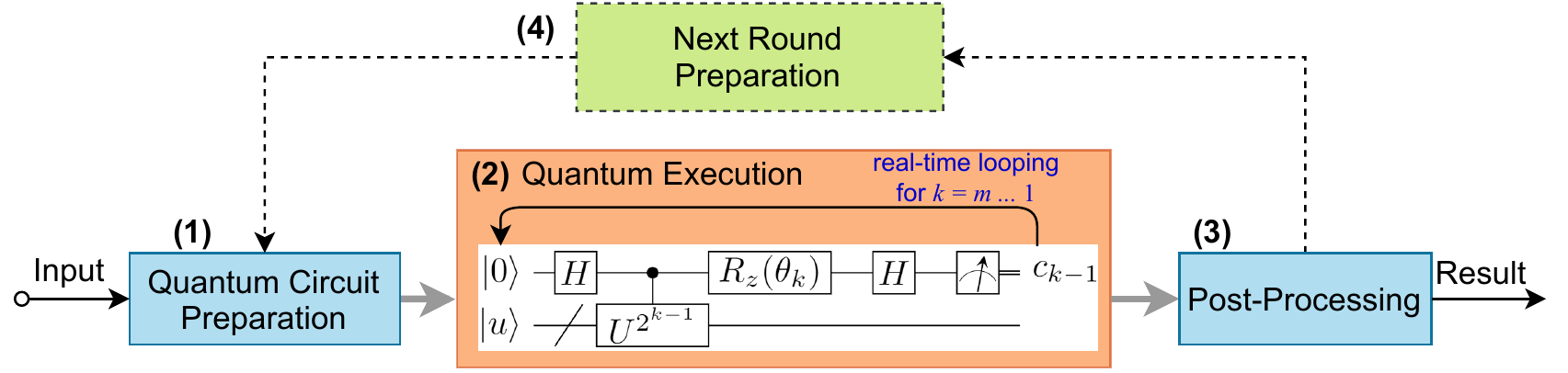}
    \caption{Flow chart of the Iterative Phase Estimation (IPE) algorithm. \ipetwo is executed by the quantum device and other steps by the classical computer.}
    \label{fig:ipe}
\end{figure}

Suppose the unitary operator $U$ has an eigenstate $\ket{u}$ with the corresponding eigenvalue $e^{i\theta}$, i.e.,
$$
U\ket{u} = e^{i\theta}\ket{u},
$$
and $\ket{u}$ can be provided, the goal of a phase estimation algorithm is to estimate the value of $\theta$.
IPE achieves this goal by utilizing only one ancillary qubit with a circuit as shown in Fig.~\ref{fig:ipe}.

To generate an $m$-bit estimation of $\theta$, a circuit consisting of $m$ sub-circuits is required with the $k$-th sub-circuit generating a one-bit measurement result $c_k$, where $k = \{m, m-1, \cdots, 1\}$. These $m$ bits together form a $m$-bit estimation to the targeting phase with the relationship $\theta = 2\pi \cdot 0.c_1c_2\ldots c_m$, where $0.c_1c_2\ldots c_m$ is binary representation of the value $\Sigma_{i=1}^{m}c_i\cdot2^{-i}$.
As shown in Fig.~\ref{fig:ipe}, each sub-circuit consists of a $z$-rotation [$R_z(\theta_k)$] and a controlled-$U^{2^{k-1}}$ operation sandwiched by two Hadamard operations ($H$) before the final measurement on the ancillary qubit.
Generally, the angle of the $z$-rotation $\theta_k$ in the $k$-th sub-circuit is determined at real-time using classical logic based on
\begin{align}
\label{eq:theta}
\theta_k = (-0.c_{k+1}c_{k+2} . . . c_m) \cdot \pi,
\end{align}
where $k\in\{1, 2, \ldots, m-1\}$, and $\theta_m =0$.
Note that the controlled-$U^{2^{k-1}}$ gate is supposed to be optimized instead of applying the controlled-$U$ gate $2^{k-1}$ times. Otherwise, the efficiency and fidelity of this algorithm will be reduced dramatically.

The flow chart of the IPE algorithm can be outlined, as illustrated in Fig.~\ref{fig:ipe}. This algorithm includes the following steps:
\begin{enumerate}[label=(\arabic*)]
    \item \label{ipestep:one} \ipeone: construct the quantum kernel program required by the following step and convert it into a hardware-readable format;
    \item \label{ipestep:two} \ipetwo: performs quantum computing. This step happens on the quantum coprocessor and comprises the following sub-steps:
    \begin{enumerate}[leftmargin=5mm, label=(\alph*)]
        \item \label{ipe:init} \kw{Initialization}: initialize the ancilla qubit into the state $\ket{0}$;
        \item \label{ipe:gates} \kw{Phase information extraction}: apply the $k$-th sub-circuit on the qubits and the phase information will be phase-kicked back to the ancilla qubit.
        \item \label{ipe:meas} \kw{Measurement}: measure the ancilla qubit;
        \item \label{ipe:rtclassical} \kw{Real-time classical computing}: compute the rotation angle $\theta_{k-1}$ using the previous measurement results based on Equation~(\ref{eq:theta}) for the next round of iteration;
        \item \kw{Iteration}: let $k\leftarrow k-1$ and repeat steps~\ref{ipe:init} -- \ref{ipe:rtclassical};
    \end{enumerate}
    \item \label{ipestep:three} \ipethree: classically calculate the desired result based on the measurement results.
    \item \label{ipestep:four} (Optional) \ipefour: calculate the new parameters based on some classical algorithm, like search;
    \item \label{ipestep:five} \ipefive: repeat steps~\ref{ipestep:one} -- \ref{ipestep:four} with the possible new parameters calculated in step~\ref{ipestep:four} in each iteration. The repetition stops after a good enough result is retrieved.
\end{enumerate}
All steps are carried out by the classical computer,  except that step~\ipetwo happens on the quantum coprocessor.

As we can see, the IPE algorithm combines classical and quantum computing in two senses.

First, \ipethree, \ipefour, and \ipeone are interleaved with  \ipetwo where slow interaction is required (indicated by thick grey lines).
In other words, the quantum state does not need to be preserved during the period when heavy classical computing happens.
In some other cases like the variational quantum eigensolver (VQE) algorithm~\cite{peruzzo2014variational}, the quantum execution needs to repeat multiple times with different quantum kernel programs, which should be calculated in step~\ref{ipestep:four} (\ipefour).

Second, to calculate $\theta_k$ at real-time and enable efficient looping over different sub-circuits in step~\ref{ipestep:two} (\ipetwo), classical instructions must be used, which calls for fast interaction with quantum operations during the quantum execution.
To accomplish the phase estimation task with one copy of the eigenstate, the IPE algorithm needs to finish execution before the qubits holding the eigenstate lose the information. Hence, it requires real-time classical instruction to enable fast looping.

A key feature of the IPE algorithm is that not all quantum gates in the circuit can be determined in step~\ref{ipestep:one} (\ipeone). The $z$-rotations in the circuit need to be calculated at real-time using classical instructions.
The feature using classical computation to decide what following operations are applied on the qubits based on previous measurement results was termed \textit{dynamic lifting} by~Green \textit{et al.}\cite{green2013quipper}.

\subsection{Execution Models of Quantum Programming Languages}
\label{sec:models}
\begin{figure}[bt]
    \centering
    \includegraphics[width=\columnwidth]{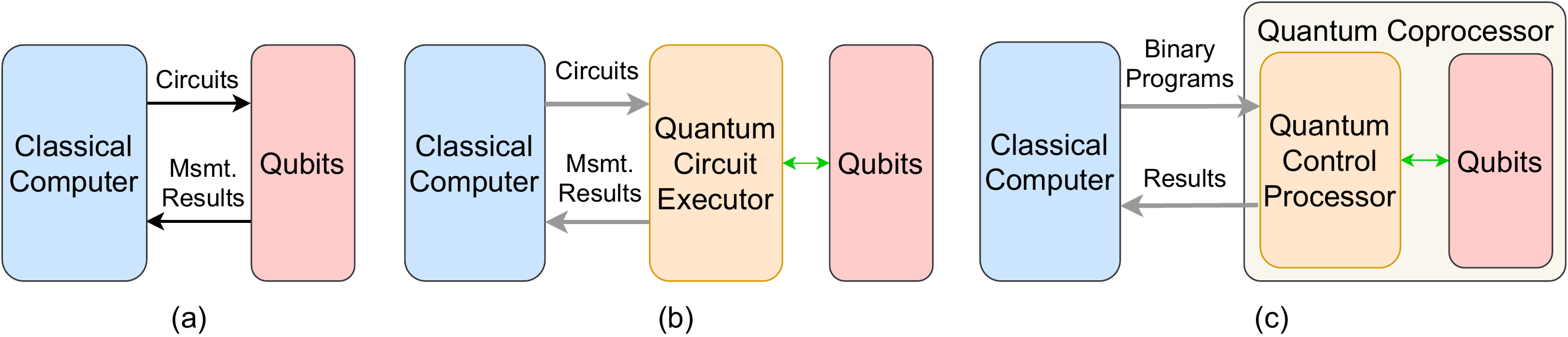}
    \caption{Various execution models for heterogeneous quantum-classical computation (HQCC). (a) Quantum Random Access Machine (QRAM) model. (b) Restricted HQCC model. (c) Refined HQCC model. While thin black lines indicate interaction between corresponding parts abstracting away hardware details, thick gray lines (thin green lines) indicate a slow (fast) interaction between corresponding parts.}
    \label{fig:models}
\end{figure}

Quantum programming languages are designed based on particular execution models, and the execution models largely define the interaction between quantum and classical computing.
By analyzing the execution or simulation of most quantum programming languages based on the circuit model,
we summarize three execution models that can depict possible interactions between classical and quantum computing resources in these languages. As shown in Fig.~\ref{fig:models}, the three models are the Quantum Random Access Machine (QRAM) model, the restricted HQCC model, and the refined HQCC model.

\subsubsection{The QRAM Model}
With the observation that practical quantum computing would take place on a classical machine with access to quantum registers, Knill proposed the Quantum Random Access Machine (QRAM) in 1996~\cite{knill1996conventions}, when there was no actual architecture for heterogeneous quantum-classical computation.
As shown in Fig.~\ref{fig:models}(a), the QRAM model consists of a classical computer and qubits or quantum registers.
The classical computer performs classical computation, controls the evolution of qubits by applying operations (or quantum circuits) on the quantum register and can read back measurement results of qubits.

The QRAM model later became a canonical model that guided the design of many quantum programming languages such as QCL~\cite{omer2003structured}, QPL~\cite{selinger2004towards}, Q Language\cite{bettelli2003toward}, Scaffold~\cite{javadiabhari2012scaffold}, \liquid~\cite{wecker2014liqui}, and \qwire~\cite{Paykin2017QWIRE}.

\subsubsection{Restricted HQCC model}
\label{sec:restricted}
The restricted HQCC model arises from the early experiment setups for quantum computing.
Except for flying qubits such as photons, stationary qubits are mostly controlled and measured via complex analog signals such as microwave or laser.
Dedicated waveform generators, modulators, and data acquisition boards are used to generate required control signals and discriminate the measurement results of qubits. These devices usually cannot execute classical instructions for classical computation as present in quantum applications.

These analog devices are usually connected to the classical computer via, e.g., Peripheral Component Interconnect Express (PCIe) bus or Ethernet.
The communication between the classical computer and the analog devices is usually of a substantial latency, which is introduced by signal transmission latency in cables, decoding the protocol in both hardware and software, memory access, operating system duties, and so on.
As observed in experiments, it could easily take milliseconds to send a command from the classical computer to the analog device, such as Tektronix Arbitrary Waveform Generator 5014~\cite{tektronixawg5000}, a control device that was widely used in controlling superconducting qubits.
Though it is possible to reduce the latency in the communication between the classical computer and the analog devices, the required engineering effort would cost more than reasonable labor and time resources.
Moreover, the achievable latency is not guaranteed because the operating system in principle has no guarantee in the response latency to an event.
Therefore, the interaction latency between classical operations executed by the classical computer and the quantum operations is very likely larger than the qubit coherence time (e.g., hundreds of microseconds for superconducting qubits~\cite{Kjaergaard2020Superconducting}).

By abstracting the quantum-classical interaction from this kind of quantum experiment setup, we get the restricted HQCC model, as shown in Fig.~\ref{fig:models}(b).
The main difference between this model and the QRAM model is the introduction of the quantum circuit executor, which can only apply a \textit{fixed quantum circuit} on qubits within the qubit coherence time. With little classical computing power, the quantum circuit executor by itself cannot support real-time feedback based on the measurement.
It is assumed that the communication between the classical computer and the quantum circuit executor has a latency comparable to or larger than the qubit coherence time.
This \textbf{slow communication} is indicated by thick grey lines in Fig.~\ref{fig:models}.
Hence, quantum-classical interaction can only happen off-line, i.e., before or after the quantum circuit is applied on qubits.

The restricted HQCC model can depict the quantum-classical interaction of some popular quantum programming languages. For example, Cirq~\cite{cirq} can only generate fixed quantum circuits without real-time feedback.
Quipper clearly distinguishes the circuit generation time and circuit execution time.
The former takes place on a classical computer and generates a fixed quantum circuit, which is executed at \textit{real-time} by the quantum processor in the latter.
Although Quipper incorporates a library to support dynamic lifting, this library is built based on the assumption that \textit{the physical quantum device has the ability to preserve qubits in long-term storage between real-time circuit invocations}~\cite{green2013quipper}. However, such kind of long-term storage is beyond the capability of NISQ technology.

\subsubsection{Refined HQCC Model}
Addressing the issue of not supporting programmable real-time feedback based on qubit measurements, dedicated quantum control architectures~\cite{ryan2017hardware, fu2017experimental, fu2019eqasm, corcoles2021exploiting} containing a control processor capable of executing interleaved quantum and classical instructions have been proposed.
The interleaved execution of quantum and classical instructions enables fast interaction between classical and quantum operations.
Hence, it can support real-time feedback as well as structural program description based on, e.g., loop and selection.
We call their execution model the \textit{refined HQCC model}, as shown in Fig.~\ref{fig:models}(c).

Compared to the restricted HQCC model, the refined HQCC model inserts a control processor between the quantum circuit executor and the classical computer.
The quantum control processor can execute quantum instructions that control the quantum circuit executor to apply quantum operations on qubits, and auxiliary classical instructions that update classical registers and direct the program flow.
Although the control processor can execute classical instructions, it can carry out very limited real-time classical computation, as NISQ qubits have limited coherence time resulting in a very limited quantum execution time.

In the refined HQCC model, the classical computer invokes the quantum coprocessor by sending binary executables instead of quantum circuits.
With similar reasons to the restricted HQCC model, the interaction between the classical computer and the quantum control processor is assumed to be slow.

This model has been adopted by heterogeneous quantum-classical programming languages like ProjectQ~\cite{steiger2018projectq}, Qiskit~\cite{qiskit}, PyQuil~\cite{pyquil}, Q\#~\cite{svore2018q}, OpenQL~\cite{khammassi2020openql}, and \sqir~\cite{hietala2021verified}~\footnote{At the time of writing, these languages have the following version numbers (if they have one): ProjectQ v0.5, Qiskit v0.27, PyQuil v2.28, OpenQL v0.9. }.
ProjectQ and Qiskit support this model because of their support for binary control (the \code{Control} meta-instruction in ProjectQ and the \code{c\_if} construct in Qiskit). Without binary control, both ProjectQ and Qiskit can only produce fixed quantum circuits in each run and could be classified into the restricted HQCC model.

\sqir~\cite{hietala2021verified} can support programmable real-time feedback based on the qubit measurements. Targeting at serving program reasoning, \sqir is too simple to support some necessary real-time classical logic  constructs and cannot support describing complex classical logic happening on the classical computer and its interaction with the quantum logic, such as the \textit{Search} step as mentioned in Section~\ref{sec:vqe_example}.
Hence,  the execution model of \sqir can be summarized as the refined HQCC model without slow communication between the classical computer and the quantum control processor.

\subsubsection{Comparison \& Discussion}
The QRAM model assumes ideal classical and quantum hardware resulting in a neat model, which enables the programmer to focus on the computational logic without worrying about implementation details.
It is possible for the classical computer to decide what following operations are applied on the qubits based on previous measurement results of qubits, i.e., to implement dynamic lifting.
However, the QRAM model may be not suitable for the NISQ technology as no constraint at all is put on the quantum-classical interaction, which is very likely beyond the capability of NISQ hardware.

The restricted HQCC model respects the latency between classical computers and qubits, which results in quantum code more suitable for execution with NISQ hardware. However, as the model does not support real-time feedback or dynamic lifting, the applications that can be described in languages based on this model are significantly limited, including the IPE example as shown in Section~\ref{sec:vqe_example}.

In contrast, the refined HQCC model can support dynamic lifting without introducing unrealistic assumptions. In the refined HQCC model, the program using dynamic lifting is compiled into instructions, and sub-circuits within each branch form an individual code block. During quantum execution, the classical instructions can direct the control flow to the corresponding code block for execution based on the qubit measurement results that are fetched at runtime.

Although a quantum language based on a simple model like the QRAM model is more appealing to the programmer for describing quantum algorithms, the refined HQCC model can enable an easier compilation from quantum algorithms described in a high-level language to NISQ hardware for execution.
We take the refined HQCC model as the underlying execution model to design the Quingo framework and the Quingo language.
After the hardware and the compilation techniques get better developed in the future, we may shift from the refined HQCC model back to the QRAM model for quantum programming.

\section{Overview of the Quingo Framework}
\label{sec:overview}

Practical quantum computing relies on the synergy between quantum and classical computing resources involving various hardware and software.
To this end, multiple quantum systems with different capabilities to support HQCC have been proposed.
However, only part of these designs or implementations is presented publicly, a big picture of integrating the heterogeneous quantum-classical software and hardware is still missing.
This requirement triggers the design of the Quingo framework.

\subsection{Design Principles}
The Quingo framework adopts the following design principles:

\begin{enumerate}[leftmargin=*, itemsep=2mm]

\item \label{prin:nisq} \kw{Matching NISQ Technology.} To ensure the framework itself and quantum applications described based on this framework can be implemented with NISQ technology, the framework should be constructed based only on technologies that have been demonstrated.

\item \label{prin:hqcc} \kw{Supporting the Refined HQCC Model.}
The Quingo framework should enable describing quantum applications in a way that naturally follows the refined HQCC model.
Classical operations that require slow communication and fast interaction with quantum operations should be easily mapped to the classical host and the control processor for execution, respectively.

\item \label{prin:modular} \kw{Modular System with a Natural Integration.}
The framework needs to be a modular system in which individual components, such as the host program, the kernel, and the compiler, are only responsible for their genuine tasks. Also, various components could be naturally integrated by some managers through clearly-defined interfaces.

\item \label{prin:exp} \kw{Quantum Experiment Support.}
The framework should support describing quantum experiments and hopefully improves the experiment efficiency.

\item \label{prin:opt}\kw{Optimization Support.}
The framework should reserve as much information as possible for the compiler to perform optimization over quantum kernels with techniques such as partial execution~\cite{javadiabhari2015scaffcc}.
\end{enumerate}

\subsection{Design Overview}
\begin{figure}[bt]
    \centering
    \includegraphics[width=\columnwidth]{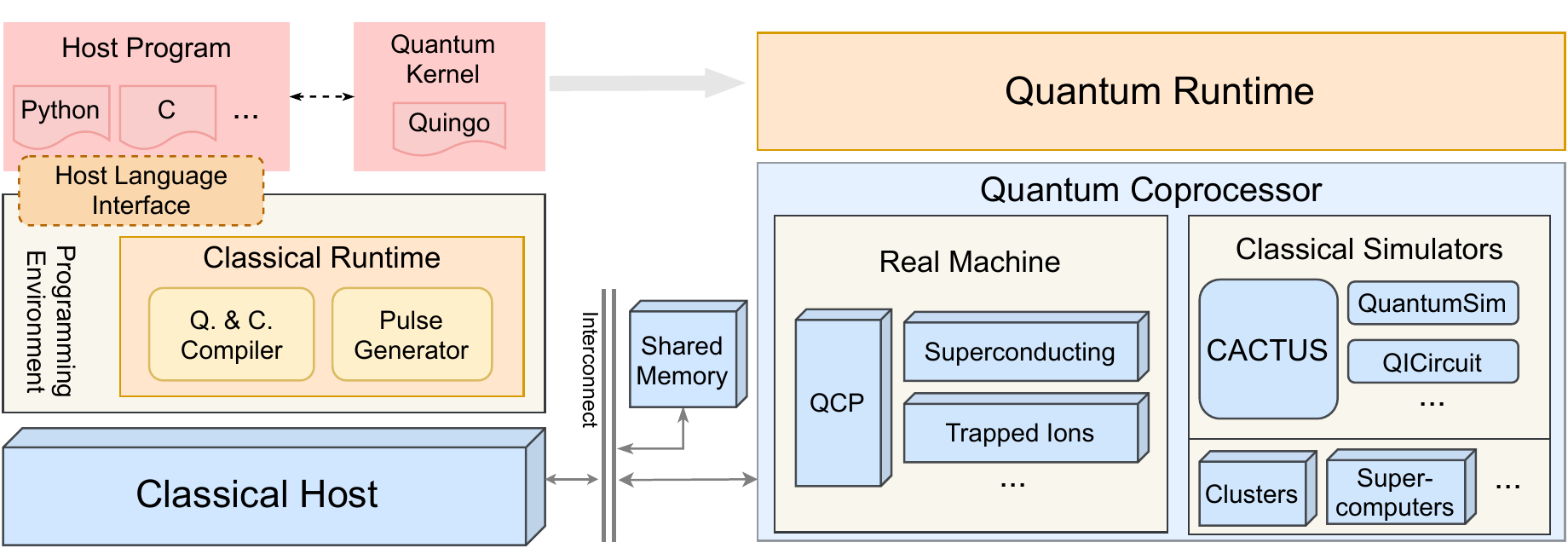}
    \caption{Overview of the Quingo framework. Except that interconnect is indicated by a pair of lines, other hardware is illustrated as cubes; software modules are presented as rounded rectangles; source files are as rectangles with a waving bottom line. Most components in the Quingo framework work on the classical host, except that the quantum kernel runs on the quantum coprocessor. The wide grey arrow indicates that the quantum kernel is compiled and uploaded to the quantum coprocessor or simulators for execution by the classical processor.}
    \label{fig:framework}
\end{figure}

Inspired by OpenCL~\cite{stone2010opencl}, an industrial standard for parallel heterogeneous computing, and guided by the design principles, we propose the Quingo framework, a proposal at the system level to integrate and manage quantum-classical software and hardware to provide the programmability over HQCC applications and experiments and map them to NISQ hardware, as shown in Fig.~\ref{fig:framework}.

The Quingo framework aims to serve quantum computing in the NISQ era based on the refined HQCC model.
It defines a minimum set of requirements on the hardware to support the refined HQCC model.
These requirements can be satisfied by nowadays hardware and have been demonstrated by previous experiments (cf. \textit{Principle~\ref{prin:nisq}}), and clarifies required software components with their responsibility boundary and interaction interface in the entire system.
Relying on the necessary software infrastructure, the framework integrates and manages the software and hardware for HQCC, helping programmers focus on describing computational logic and map the quantum applications to the NISQ hardware.
The Quingo framework comprises the following key components:

\begin{itemize}[leftmargin=*, itemsep=2mm]

\item \textbf{Quantum programs}: The quantum program is described in two parts: the classical host program and the quantum kernel;

\item \textbf{Compilers}: At least two compilers should be used, including a classical compiler or interpreter, and a quantum compiler. Besides that, a pulse generator is required to generate pulses for customized operations in quantum experiments or quantum optimal control~\cite{werschnik2007quantum, leung17speedup, shi2019optimized}.

\item \textbf{Hardware}: The Quingo hardware platform comprises a classical host, a quantum coprocessor, and a shared memory  accessible to both the host and the coprocessor. The quantum coprocessor includes a control processor capable of executing interleaved quantum-classical instructions to control qubits.
Note, the quantum coprocessor can be simulated using simulators such as CACTUS~\cite{fu2020cactus} with QuantumSim~\cite{brien2017density} or QICircuit~\cite{guo2019genreal}.

\item \textbf{Runtime system}:
A runtime system serving as the infrastructure of this framework that integrates and manages various quantum and classical software and hardware components.
\end{itemize}

To match the HQCC model, quantum applications based on the Quingo framework are described in two parts: the host program described in a classical language such as Python or C, and the quantum kernel described in Quingo.
For example, the IPE algorithm as shown in Section~\ref{sec:vqe_example} is described jointly by a Python host program (Code~\ref{code:ipe_host}) and a Quingo kernel (Code~\ref{code:ipe_kernel}) (Section~\ref{sec:host} and Section~\ref{sec:language} detail the host and kernel languages, respectively).
The programmer is responsible for putting all classical computation that requires fast interaction with qubits in the kernel, and the other classical tasks in the classical host program. Since the host program is fully-fledged, any classical computing that does not require fast interaction is suggested to be offloaded to the host program in a classical language, which can significantly ease the burden in developing classical libraries in Quingo.
This follows \textit{Principle~\ref{prin:hqcc}}.

Ideally, the host program should focus on describing the classical computing task and invoking the quantum kernel;
the quantum kernel focuses on describing the computing task running on the quantum coprocessor;
and the (quantum) compiler focuses on compiling the given Quingo program with given parameters.
To get a seamlessly workable system, the framework should (i) provide an interface for the host language to call the quantum kernel with parameters, (ii) trigger the quantum compiler to compile the kernel, (iii) load the quantum binary to the quantum coprocessor and trigger its execution, and (iv) read the kernel execution result and return it to the host language.
To this end, the Quingo runtime system is introduced.
Besides that, the selection and configuration of the target backend are also done by the runtime system, instead of by the host program.
In this way, the same quantum program can be executed by real hardware or simulators without any modification in the source files of the quantum program (cf. \textit{Principle~\ref{prin:modular}}). The runtime system is introduced in Section~\ref{sec:rtsys}.

A key component of the Quingo framework is the language used to describe the quantum kernel.
The kernel language should support the usage of opaque operations and controlling the timing control to describe quantum experiments (cf. \textit{Principle~\ref{prin:exp}}).
As describing real-time classical logic using embedded DSL would result in complicated code, as illustrated in Code~\ref{code:pyquil_loop}, the quantum language is suggested to be an external DSL (cf. \textit{Principle~\ref{prin:hqcc}}). We have designed the Quingo quantum programming language to satisfy these requirements, which is introduced in Section~\ref{sec:language}.

Previous work has shown that quantum algorithms can be significantly optimized with partial execution~\cite{green2013quipper, javadiabhari2015scaffcc} after recognizing the different phases of a quantum program.
Based on the refined HQCC model, the Quingo framework refines previous life-cycle models into a six-phase one. Considering the capability of the control processor to execute classical instructions, Quingo delays the generation of quantum code until the point immediately before quantum state evolution starts to provide as much information as possible to the quantum compiler (cf. \textit{Principle~\ref{prin:opt}}).

In the rest of the paper, key techniques in the Quingo framework are introduced in the following section and the Quingo language in Section~\ref{sec:language}.
\section{Key Techniques in the Quingo Framework}
\label{sec:hqcc}

We first introduce the quantum program life-cycle model in Section~\ref{sec:six-phase}, which defines the routine of how the entire framework works based on the refined HQCC model. It divides the entire life of a quantum program into six phases and defines the responsibility of each component or the programmer in each phase.
The software system of the framework, which is embodied in the runtime system, is introduced in Section~\ref{sec:rtsys}.
Section~\ref{sec:host} presents the requirements and constraints that the Quingo framework put on the host language and how the host language interacts with the quantum kernel via the provided programming interface.
The Quingo framework also puts some requirements on the compilation of the quantum kernel, to maximally utilize the optimization space as reserved by the six-phase life-cycle model. This is introduced in Section~\ref{sec:jit}.
Last but not least, data exchange among different components in the framework should be supported by both the software and hardware. We delay the introduction of the mechanism to exchange data after we introduce the Quingo language in Section~\ref{sec:quingo_data_ex}.

\subsection{Quantum Program Life Cycle}
\label{sec:six-phase}

A quantum program life-cycle model can help clarify what task should be carried out by which component in what order, which can help the programmer to understand the implication of the code they write, and also guide a seamless integration of various quantum and classical components into a working system.
The design of the quantum program life-cycle model aims to reserve as big optimization space over quantum programs as possible and support quantum-classical interaction with timing constraints satisfied.
Based on the refined HQCC model, we propose a six-phase quantum program life-cycle model, of which the feasibility has been demonstrated by the common workflow as observed in nowadays quantum experiments and quantum algorithms.
It includes six phases:
\begin{enumerate}[label=(\arabic*)]
    \item \textbf{Editing}: Quantum programmers describe the quantum application using a classical programming language (such as Python or C) for the host program and Quingo for the quantum kernel. The required configuration (such as what primitive operations can be used by the hardware) is also provided at this stage.

    \item \textbf{Classical compilation} (optional):  A conventional compiler, such as GCC, compiles the host program and outputs a classical binary. Note that this phase may  not be needed for interpreted languages such as Python.

    \item \label{phase:pre_cl_exe}\textbf{Classical pre-execution}: The classical host executes the classical binary up to the moment calling the quantum function defined in the kernel. At this moment, all parameters used to invoke the quantum function (called the \textit{kernel interface parameters})  have been determined.

    \item \label{phase:quantum_compile}\textbf{Quantum compilation}: The quantum compiler compiles the kernel into a quantum binary consisting of quantum-classical mixed instructions with possible extra data. At this step, the quantum compiler can use the kernel interface parameters passed in to perform optimization by, e.g., partial execution.

    \item \textbf{Quantum execution}: The control processor loads and executes the quantum binary. According to the executed instructions, the control processor updates classical registers, performs flow control, and applies corresponding quantum operations over qubits. In this way, the quantum state evolves under program control, accomplishing the kernel computational task. Then, the computation result is sent to the classical host by writing to the shared memory between the quantum coprocessor and the host.
    \item \label{phase:post_cl_exe}\textbf{Classical post-execution}: The classical binary continues execution, reads the quantum computation result, and performs possible post-processing.

\end{enumerate}
When required as in algorithms like VQE, phases~\ref{phase:pre_cl_exe} to \ref{phase:post_cl_exe} (from classical pre-execution to classical post-execution) could be repeated multiple times to reach a good enough result.

Note that in some scenarios, the quantum compilation phase or part of it can be brought forward to happen at the same time as the classical compilation. For example, when the execution of the quantum kernel does not depend on the classical parameters, and deep optimization over the quantum algorithm is not so critical compared to the requirement to execute the quantum circuit as soon as possible, such as programs used for quantum communication such as teleportation~\cite{bennett1993teleporting}. Naturally, the semi-static compilation, as described in Section~\ref{sec:jit}, might be disabled, and the quantum binary generated without much optimization is loaded to the quantum control processor awaiting execution.

\begin{lstfloat}[bt]
\begin{lstlisting}[
style=mypython,
numbers=left,
linewidth=0.94\columnwidth,
xleftmargin=0.06\columnwidth,
basicstyle=\linespread{1.0}\scriptsize\ttfamily,
label=code:ipe_host,
caption=Python host of the IPE algorithm
]
# host.py: the host of the iterative phase estimation algorithm
from qgrtsys import if_quingo
''' Call the Quingo kernel to estimate the oracle phase.
    Repeat $n$ times, each time receive $m$ bits.
'''
def ipe(m: int, n: int) -> float:
    res = 0
    for i in range(n):
        if not if_quingo.call_quingo("ipe.qu", "ipe", m):
            raise SystemError("The execution of the quantum kernel fails.")
        res += if_quingo.read_result()
    return res / n

\end{lstlisting}

\begin{lstlisting}[
  style=QuingoStyle,
  numbers=left,
  linewidth=0.94\columnwidth,
  xleftmargin=0.06\columnwidth,
  basicstyle=\linespread{1}\scriptsize\ttfamily,
  caption=Quingo kernel of the IPE algorithm.,
  label=code:ipe_kernel,
]
import operations.*
import config.json.*

// kernel.qu: Iterative phase estimation algorithm.
// Input: the number of bits of the estimation
// Output: the estimation result
operation ipe(m: int) : double {
    double theta = 0.0;      // = theta_k / PI
 
    using (ancilla: qubit, eigenstate: qubit) { // Allocate two qubits
        if (!measure(eigenstate)) {             // prepare the eigenstate |1>
            X(eigenstate, PI);   
        }
        for (int k = m - 1; k >= 0; k -= 1) {
            init(ancilla);                      // reset ancilla to |0>
            H(ancilla);
            
            control(ancilla, oracle(eigenstate, k));  // controlled-U^(2^i)
            Z(ancilla, -PI * theta);
            H(ancilla);

            if (measure(ancilla)) {             // Update the estimated phase
                theta = theta / 2.0 + 0.5;
            } else {
                theta /= 2.0;
            }
        }
    }
    return PI * theta;
}
\end{lstlisting}

\end{lstfloat}

We take the IPE and VQE algorithm as examples to illustrate the match between the six-phase life-cycle model and quantum algorithms.
The IPE algorithm is first described in a classical language and a quantum language (see Code~\ref{code:ipe_host} and Code~\ref{code:ipe_kernel}), this corresponds to phase (1) (\kw{editing}).
Being an interpreted language program, the Python host in Code~\ref{code:ipe_host} can start execution without compilation (omitting the optional phase (2): \kw{classical compilation}).
The classical host program starts execution in step~\ref{ipestep:one} (\kw{quantum kernel preparation}) and determines all parameters required by the quantum kernel (phase (3): \textit{classical pre-execution}).
Thereafter, the quantum compiler compiles the quantum kernel during which the kernel can be optimized (phase (4): \textit{quantum compilation}).
The quantum kernel starts execution utilizing quantum and classical operations with fast interaction in between in step (2) (\kw{quantum execution}) and finishes with the measurement results generated (phase (5): \kw{quantum execution}).
The host program fetches the measured data and calculates the required result in step (3) (\kw{post-processing}).
For VQE-like algorithms, multiple iterations of the quantum execution with different parameters are required.
Based on the previously calculated result, the classical computer searches a new set of parameters in step (4) (\kw{search}). Steps (3) and (4) (\kw{post-processing} and \kw{search}) together form phase (6) (\textit{classical post-execution}).
The newly-generated parameters is then employed to prepare the quantum kernel to be used in the next round, which restarts the routine from phase (3) (\textit{classical pre-execution}).

\subsection{Runtime System}
\label{sec:rtsys}

Since the host program, the kernel program, the compiler, and the quantum control processor in the framework are expected to be only responsible for their own tasks,
they cannot directly work in collaboration due to the interaction among them. The following problems need to be resolved:
\begin{itemize}
    \item Since the host program is not responsible for compiling the quantum program, when and which component should trigger the quantum compiler to compile the quantum program?

    \item The kernel interface parameters provided by the host program reside in the memory space corresponding to the process of the host program on the host machine, and cannot be directly read by the quantum compiler working in another process on the host machine. How to pass the kernel interface parameters to the quantum compiler?

    \item Which component should upload the quantum code to the quantum control processor and trigger the execution after its generation?

    \item How to pass the kernel execution result back to the host program?
\end{itemize}

\begin{figure}[bt]
    \centering
    \includegraphics[width=0.9\columnwidth]{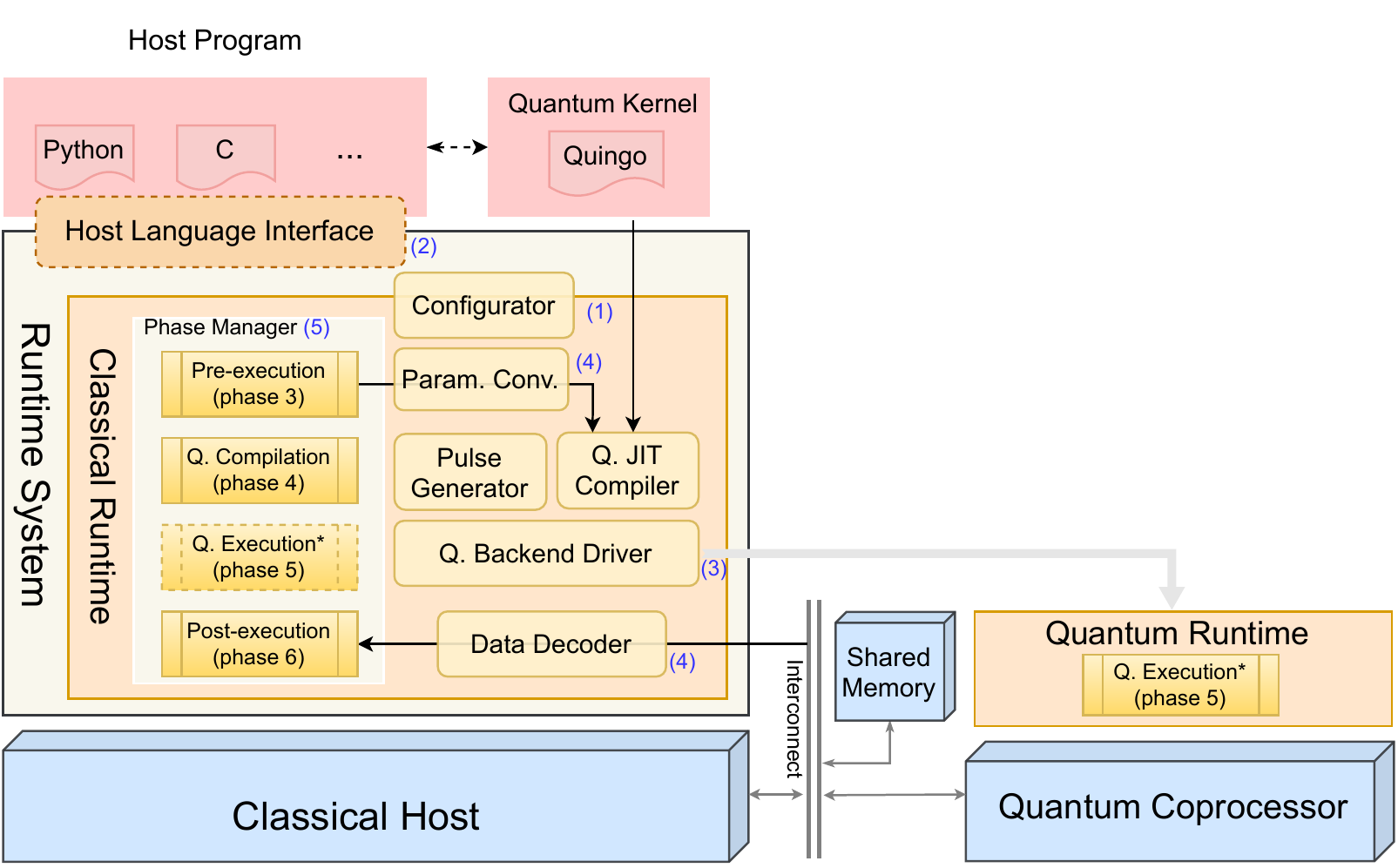}
    \caption{Quingo runtime system supporting the six-phase quantum program life-cycle model. The number of each part of the runtime system is labeled in purple on the right side of the corresponding module. The \textit{Q. Execution} phase in the classical runtime is framed with dashed lines, which means that the work is actually done by the coprocessor in the quantum runtime.}
    \label{fig:rtsys}
\end{figure}

To get a full HQCC system with components working together seamlessly, we propose the Quingo runtime system as the supportive environment of the Quingo framework, as shown in Fig.~\ref{fig:rtsys}.
The Quingo runtime system is designed as a library or daemon running on the classical host machine, which provides an application programming interface (API) to the classical host language and manages the interaction among the classical host and the quantum kernel at both software and hardware level.
It mainly consists of five parts:
\begin{enumerate}
    \item A system configurator, which is in charge of configuring the execution environment for the quantum program;
    \item A host language interface, which enables the host program to call quantum kernels to utilize the quantum coprocessor to solve problems and read the result from the quantum kernel;
    \item An interface to call various quantum backends to execute the quantum code and enable them to return the computation results. This interface is implemented as various quantum backend drivers;
    \item A parameter converter and kernel result decoder, which are responsible for enabling the communication between the host and the kernel;
    \item A phase manager, which is responsible for triggering corresponding activities at different phases of the program life-cycle model.
\end{enumerate}

The runtime system supports the execution of quantum programs as described in the following. Before executing the quantum program, the programmer configures the execution environment through the configurator. For example, selecting a real machine or a simulator to execute the quantum program is done at this stage. When the host program calls the quantum kernel through the host language interface (see Section~\ref{sec:host}), the runtime system is activated, which delivers the control to the phase manager. This moment corresponds to a time point in  phase~(3) (\kw{classical pre-execution}). The phase manager then passes the parameter to the Quingo compiler (see Section~\ref{sec:read_data}), and triggers the compiler and the pulse generator (phase~(4): \kw{quantum compilation}). After the compiler returns the quantum code, the phase manager uploads the quantum code to the shared memory and triggers the quantum coprocessor to execute it (phase~(5): (\kw{quantum execution})). The phase manager then waits for the quantum kernel to return. The quantum kernel writes the execution result to a piece of the shared memory with a starting address previously defined. Thereafter, the phase manager triggers the data decoder to decode the returned result in the shared memory and returns it to the host program in a format readable for the host language for post-execution (phase~(6): \kw{classical post-execution}). The data decoder is introduced in Section~\ref{sec:quingo_data_ex}. After that, the phase manager exits and returns the control to the host program.

All work done by the runtime system is transparent to the host program, the Quingo kernel, and the Quingo compiler. As a result, they can focus on their genuine roles without worrying about other details. For example, the host program only needs to describe the classical logic processing the parameters sent to and the results received from the Quingo kernel. The host program needs not learn about the concrete hardware architecture or the control methods over the quantum coprocessor. Also, both the host program and the Quingo kernel can directly use the parameters or results passing in  between without touching the communication details. We believe that this design helps to ease the development of HQCC applications.

\subsection{Requirements on the Host Program}
\label{sec:host}
% -----------------------------------------------------------------------------------
In the editing phase, the programmer is responsible for describing not only the kernel in Quingo, but also the host program.
The Quingo framework allows any general-purpose classical language to be used as the host language, such as Python.
The Quingo runtime system provides a set of APIs, currently in Python, for calling the Quingo kernel and configuring the execution environment.
A classical language implementing the APIs provided by the runtime system can be an eligible host language in the Quingo framework.
It offers the freedom for the programmer to choose a suitable classical language according to his/her requirements.

The host program of the IPE algorithm is shown as an example in Code~\ref{code:ipe_host}.
With the interface provided by the Quingo runtime system (\pyth{qgrtsys}), the host program calls the quantum kernel through the interface function \pyth{if_quingo.call_quingo} at line~8. The quantum function \quingo{ipe} defined in the file \textit{kernel.qu} is called with two kernel interface parameters \pyth{m} and \pyth{n}, which are both integers in this algorithm.
After the kernel finishes execution successfully, the host retrieves the  kernel execution result using the function \pyth{if_quingo.read_result} for post-classical processing.
\lines{8}{11} in Code~\ref{code:ipe_host} repeatedly call the Quingo kernel for \code{n} times, which is followed by a simple averaging at line~12. The actions behind this function correspond to phases~(3) -- (6) (\kw{classical pre-execution} to \kw{classical post-execution}) in the six-phase life-cycle model.

Although the host program is allowed to call quantum kernels any number of times at any point, the HQCC model assumes the quantum state will \textbf{not} be preserved between two consecutive calls to the quantum kernel.
If some classical computation is required to interact with quantum operations in real-time, the programmer should put these classical operations in the kernel.
In this way, the timing of quantum-classical interaction can be assured.

Note, except for the methods and APIs defined by the runtime system used to call the Quingo kernel and retrieve the quantum kernel result, there are no other constraints on the host program written in the classical language.
Hence, the host program can import and use any packages or libraries available in the classical language.
Since classical operations requiring fast interaction with quantum operations are limited, and classical computation that only requires slow communication with qubits can be offloaded to the host program and described using fully-fledged classical programming languages, the Quingo language is not required to develop libraries for complex classical computations to enhance its expressiveness for quantum
applications.

\subsection{Requirements on the Quantum Compiler}
\label{sec:jit}
Each time the host program calls the quantum kernel with a set of determined parameters, the quantum compiler is called subsequently to translate and optimize the quantum kernel and generate the quantum code for execution.
The quantum compiler also forms a critical component of the Quingo framework.
This section introduces the requirements for the compiler of the Quingo framework. We will not discuss a concrete quantum compiler design as it is not part of the framework.

\subsubsection{Intermediate Representation}
Since the refined HQCC model allows the execution of interleaved quantum and classical instructions, the quantum compiler should be capable of representing both quantum and classical computational logic to generate quantum-classical mixed binaries.
Many existing quantum compilers adopt some format of quantum circuits as the intermediate representation (IR) of quantum applications.
For example, the internal representation (IR) of Qiskit is a directed acyclic graph with quantum-operation nodes representing  a quantum circuit. Quantum-circuit-based IR can be cumbersome when used to represent classical constructs in quantum algorithms.

Taking statements as the basic element of IR can enable a flexible description of HQCC applications with rich quantum-classical interaction~\cite{svore2018q}.
As a consequence, the quantum compiler is best constructed with an IR based on statements or a hierarchical structure such as MLIR~\cite{lattner2020mlir}, instead of quantum circuits, to enable easy manipulation.
In this way, the classical constructs or operations in the quantum kernel can be translated into classical instructions mixed with quantum instructions, which are eventually executed by the control processor.

Since manipulating the timing of operations is crucial for quantum experiments and presents increasing importance in optimization over NISQ applications~\cite{corcoles2021exploiting}, the IR should also support representation and manipulation of the timing of operations.

\subsubsection{Semi-Static Optimization}

Although it is possible to describe classical logic in the quantum programming language, it does not mean that all classical logics described in the kernel should be translated into instructions executed on the control processor.

More information provided to the compiler usually helps better optimization.
As the kernel interface parameters are fully determined before calling the kernel, it can also enlarge the optimization space for the quantum compiler.
Though not necessary, the Quingo framework suggests that the quantum compiler highly utilizes classical optimization techniques, such as procedure cloning, constant propagation, dead code elimination~\cite{javadiabhari2015scaffcc}. In addition, partial execution is also an important technique that should be  harnessed to resolve and eliminate classical operations in the quantum program.

For example, if the variable \quingo{x} in the branch statement \lstinline[style=StyleQuingoinline]!if(x) {X(q);} else {Y(q);}! is a kernel interface parameter, the compiler can know the actual value of \quingo{x} and then perform partial execution during compilation.
Hence, the branch structure can be eliminated and the compiled result of this branch statement would be a simple quantum assembly instruction \quingo{X q} or \quingo{Y q}.
By doing so, the classical instructions can be further reduced, which better matches the NISQ hardware where the classical computing power on the control processor is limited.

\subsubsection{Rethinking about Quantum Compilation}
In the Quingo framework, the quantum code is not generated when the host program starts execution.
Quantum compilation happens after the classical host calls the quantum kernel, which is the compile time for the quantum kernel but the runtime for the classical host.
We call quantum compilation happening at this stage \textit{semi-static compilation}.
By delaying the generation of quantum code to the last moment, i.e., just before its execution, the Quingo framework maximizes the optimization space reserved for the quantum compiler.
To make it clear, we term phases~(3) -- (6) as the classical runtime, and phase~(5) as the quantum runtime. Quantum state evolution only happens during the quantum runtime.

Note that, when applicable, optimization techniques in classical Just-In-Time (JIT) compilers could also be adopted by the quantum semi-static compiler, such as collecting statistics about how the program is actually running to rearrange and recompile for an optimal performance.

\section{The Quingo Language}
\label{sec:language}

The Quingo framework suggests an external DSL used to describe the quantum kernel. In this section, we present the Quingo language, which is designed following the requirements of the Quingo framework based on the refined HQCC model.
The design of the Quingo language is guided by the following principles:
\begin{enumerate}[leftmargin=*, itemsep=2mm]
\item \label{prin:neat_syntax}\kw{Intuitive and Concise Syntax.}
The syntax should be intuitive to reduce the barrier for new users in learning this language. It should have native comprehensive support for structured programming and other commonly-used program patterns to enable a concise description of HQCC algorithms.

\item \label{prin:fast_interaction}\kw{Native Support for Fast Interaction.} Classical operations and program flow control on the control processor should be naturally described without relying on dedicated variables or program structures.

\item \label{prin:choice}\kw{Minimal Design Concepts.}
The Quingo language aims at providing core features to support HQCC with a minimal set of concepts. Features that can be implemented using the core features will be provided by libraries. The current design of Quingo should not impede its extension to support high-level programming features in the future, such as adding control qubits to quantum operations or inverting operations.
\end{enumerate}

The Quingo language adopts a two-level design to define the language to ensure the extensibility of the language and enable a practical engineering implementation.
At the core level, the syntax is defined with a minimal set of concepts, data types, and rules which aim to be expressive and comprehensive enough.
At the user level, syntactic sugar is constructed based on the core syntax to improve programming efficiency.
The benefit of this method is double-folded.
First, the core syntax being minimal enables a relatively easy and formal definition of this language and simplifies the compiler design and formalization.
Second, syntactic sugar can be added by modifying the compiler front-end without affecting the core syntax and intermediate representation, which enhances the extensibility of the language while at the same time keeps the language steady.
This section gives a detailed description of the Quingo language. We focus on the core syntax since the user-level syntax is still under development.

The Quingo language supports standard statements and expressions available in conventional imperative languages, as well as special syntax for describing quantum algorithms and experiments.
The file \href{https://github.com/quingo/compiler_xtext/blob/master/docs/syntax_core.md}{syntax\_core.md}\footnote{Available at:
\textcolor{blue}{
\url{https://github.com/quingo/compiler_xtext/blob/master/docs/syntax_core.md}}.} presents the core syntax of this language in the Backus-Naur Form (BNF) format.
The \quingo{import} and \quingo{package} statements make up a simple module system (Section~\ref{subsec:lang-modu}).
The Quingo language has a strong static type system where each variable is explicitly defined with a type (Section~\ref{subsec:lang-types}). The \quingo{using} statement is for the allocation and de-allocation of qubits (Section~\ref{subsubsec:qubit}). General control-flow structures are supported, including \quingo{if}-\quingo{else}, \quingo{while}, \quingo{break}, \quingo{continue}, and \quingo{return}. Functions in the Quingo language are called operations. There are two kinds of operations, \quingo{opaque} and \quingo{operation} (Section~\ref{subsec:lang-ops}). An operation call can be associated with timing constraints to control the execution time of the operation (Section~\ref{subsec:lang-timing}).

For demonstration purposes, we use the program  shown in Code~\ref{code:ipe_kernel} as a running example throughout this section. Code~\ref{code:ipe_kernel} implements the IPE kernel to estimate the eigenvalue $e^{i\theta/2}$ of the given oracle $R_z(\theta)$ where $\theta$ is unknown. As the oracle operates only one qubit, the IPE kernel can be implemented using only two qubits.

\subsection{Module System}
\label{subsec:lang-modu}
A module system is used to organize program code in the Quingo language. Multiple operations (explained later in Section~\ref{subsec:lang-ops}) related to a common topic can be collected into a \textit{package} to enable separate compilation, avoid name conflict, and ease code distribution. A package is declared with the \quingo{package} statement at the top of the source file, indicating that all operations defined in this file belong to this package. Code~\ref{code:vqe-ops} shows a code package named \texttt{operations}, in which a bunch of opaque operations are defined. Operations inside a package can be imported by other files with the \quingo{import} statement, as shown in \lines{1}{2} in Code~\ref{code:ipe_kernel}.

\begin{lstfloat}
\begin{lstlisting}[
  style=QuingoStyle,
  numbers=left,
  linewidth=0.94\columnwidth,
  xleftmargin=0.06\columnwidth,
  basicstyle=\linespread{1}\scriptsize\ttfamily,
  caption=A package containing opaque operations imported by the VQE kernel,
  label=code:vqe-ops,
]
package operations

opaque I(q:qubit): unit;
opaque H(q:qubit): unit;
opaque X(q: qubit, angle: double): unit;
opaque Y(q: qubit, angle: double): unit;
opaque Z(q: qubit, angle: double): unit;
opaque CNOT(ctrl: qubit, target: qubit): unit;
opaque measure(c:qubit): bool;
\end{lstlisting}
\end{lstfloat}

\subsection{Type System}
\label{subsec:lang-types}
The Quingo language has a strong static type system.
Types in the Quingo language include primitive types for classical and quantum data, operation types, and composite types.
Besides these, the Quingo language has two special types dedicated to timing control discussed in Section~\ref{subsec:lang-timing}.

\subsubsection{Primitive Classical Types}
Considering the limited classical computational power of the control processor, the Quingo language only allows four primitive classical types, i.e., \quingo{bool}, \quingo{int}, \quingo{double}, and \quingo{unit}. The \quingo{unit} type is merely used to describe the return type of an operation that has no return value. For other classical types, basic arithmetic and logical operations are supported.

\subsubsection{Primitive Quantum Type}\label{subsubsec:qubit}
The Quingo language defines \quingo{qubit} type as the only primitive type for quantum data.
One or more qubits can be allocated from a pool with the \quingo{using} statement (e.g., line~10 in Code~\ref{code:ipe_kernel}), and later be referenced using \quingo{qubit}-type variables (e.g., the variables \quingo{ancilla} and \quingo{eigenstate} in line~10 in Code~\ref{code:ipe_kernel}).
Scope of a \quingo{qubit}-type variable is within the block (between the curly braces) corresponding to the \quingo{using} statement.
Qubits are allocated at the beginning of the \quingo{using} block and automatically de-allocated (i.e., returned to the pool) when exiting this block.
The automatic de-allocation semantics avoids leakage of the qubit resource. Qubits can only be manipulated by quantum operations  (\lines{11}{21} in Code~\ref{code:ipe_kernel}).
The only way to read out information from a qubit is using a measurement operation, which projects the qubit to the computational basis state and returns the measurement result as a \quingo{bool} value (lines~11 and 21 in Code~\ref{code:ipe_kernel}).

Unlike other quantum programming languages, the Quingo language assumes the qubit is in an \emph{unknown} state instead of $\ket{0}$ upon allocation, and the programmer is responsible for initializing it.
This assumption is necessary to quantum experiments where the initialization has to be explicit.

\subsubsection{Operation Type}
The type of an operation consists of a parameter type and a return type. For instance, \quingo{qubit->unit} is the type of an operation on a single qubit that returns nothing.
Parameter type of a multi-parameter operation would be a tuple,
e.g., the type of the \code{oracle} operation (line~18 in Code~\ref{code:ipe_kernel}) is denoted as
    \quingo{(qubit,int)->unit}.

\subsubsection{Composite Types}
The Quingo language supports using \textit{array} and \textit{tuple} to define data collections. Any valid Quingo type can be the element type of \textit{array} and \textit{tuple}.

An array is an ordered sequence of elements of the same type, i.e., \textit{array} is a homogeneous collection type. Arrays can be modified dynamically by inserting, deleting, or replacing values. Quingo arrays are jagged arrays, i.e., sub-arrays can have different lengths.

A tuple is an ordered, heterogeneous collection of elements.
In the Quingo language, tuples are immutable and mostly used to pass parameters to and return values from operation calls.

\subsection{Operations}
\label{subsec:lang-ops}
The Quingo language supports functions for structured programming. Functions are called \textit{operation}s to emphasize that they are processes performing quantum operations on qubits for a particular purpose. There are two kinds of operations in the Quingo language, opaque operations and user-defined operations, defined with keywords \quingo{opaque} and \quingo{operation}, respectively.

\subsubsection{Opaque Operation}
In theory, there exists a set of universal quantum gates that can approximate any other quantum gates with arbitrary precision albeit at the cost of longer operation sequences using some decomposition techniques, such as repeat-until-success~\cite{paetznick2014repeat}.
The universal gate set is not unique,
and different quantum technologies may utilize a different primitive gate set considering the implementation difficulty.
The Quingo language does not define any built-in quantum primitives. Instead, a mechanism is provided
to declare platform-dependent primitive operations. The mechanism consists of two parts: (1) a platform-dependent configuration file that describes the available primitive operations, (2) opaque operation declaration statements defining the interface for these primitive operations. The second part has already been shown in Code~\ref{code:vqe-ops}. Code~\ref{code:vqe-config} shows part of the configuration file imported by the IPE kernel (Code~\ref{code:ipe_kernel}).

The format of the configuration file is quite similar to JSON, with a leading \quingo{package} statement declaring the package name. The configuration file consists of two sections, a platform definition section (\lines{3}{11}) and an operation definition section (\lines{12}{53}). The former describes features of the target architecture, e.g., the number of available qubits, and the latter provides information about primitive operations on that architecture.

\begin{lstfloat}
\begin{lstlisting}[
  style=mypython,
  numbers=left,
  linewidth=0.94\columnwidth,
  xleftmargin=0.06\columnwidth,
  basicstyle=\linespread{1}\scriptsize\ttfamily,
  caption=Platform-dependent configuration file for Quingo,
  label=code:vqe-config,
]
package config.json

platform_def = {
    "num_qubits": 5,
    "single_qubit_gate_fidelity": {
        "xy_rotations": [0.997, 0.992, 0.994, 0.996, 0.995]
        "z_rotation": [0.985, 0.989, 0.990, 0.984, 0.977]
    },
    "qubit_coupling" : [[0, 1], [1, 2], [2, 3], [3, 4], [4, 5]],
    "two_qubit_gate_fidelity": [0.985, 0.989, 0.990, 0.984, 0.977]
}
op_def = {
    "X": {
        "duration": 20e-9,    "num_qubits": 1,
        "params": [{"name": "theta", "type": "double"}],
        "semantics": {
            "type": "rotation",
            "rot_axis": [1, 0, 0],  # x-axis
            "rot_angle": "theta"
        }
    },
    "Y": {
        "duration": 20e-9,    "num_qubits": 1,
        "semantics": {
            "type": "pulse",
            "assembly": {"type": "eqasm", "name": "y"},
            "pusle": {
                "pulse_name": "gaussian",
                "params": {
                    "amplitude": 0.36,
                    "sigma": 20e-9,
                    "length": 4,
                    "sample_rate": 1e9,
                    "phase": PI/2
                }
            }
        }
    },
    "H": {
        "duration": 40e-9,    "num_qubits": 1,
        "semantics": {
            "type": "matrix",
            "matrix": [ [[0.707107,0.0], [0.707107,0.0]],
                         [0.707107,0.0], [-0.707107,0.0]] ]
        }
    },
    "measure": {
        "duration": 600e-9,
        "semantics": { "type": "measure", "assembly":"MeasZ", "return": bool }
    }
    # more operation declarations
    # ......
}
\end{lstlisting}
\end{lstfloat}

An operation is defined with a few properties. The \textit{duration} and \textit{num\_qubits} properties represent the duration the operation lasts and the number of target qubits, respectively. The Quingo language supports the definition of parametric operations. For instance, the \quingo{X} operation in Code~\ref{code:vqe-config} is defined as rotation along the $x$-axis with the rotation angle specified by the parameter \texttt{theta} (\lines{16}{20}). The \textit{semantics} property describes the operation's inherent quantum semantics, that is, how this operation transforms the target qubit(s) state. Code~\ref{code:vqe-config} presents four different ways to define the \textit{semantics} property:
\begin{enumerate}
\item A \textit{rotation} along a specific axis by a particular angle, such as the \quingo{X} operation at \lines{16}{20}.
\item A \textit{pulse} with the corresponding assembly code, such as the \quingo{Y} operation at \lines{24}{37}.
\item A unitary \textit{matrix} describing the transformation on the state vector when applying this operation on the target qubit(s), such as the \quingo{H} operation at \lines{41}{45};
\item The \textit{measure} semantics dedicated to the \quingo{measure} operation (\lines{47}{50}).
\end{enumerate}

To support features exposed by the Quingo language, the IR of the Quingo compiler should be capable of representing the \textit{matrix}, \textit{rotation}, and \textit{measure} semantics and can perform code analysis and transformation based on this information. Operations with \textit{pulse} semantics are treated as black boxes during the analysis and transformation phases. The associated information is only used by the compiler backend for code generation.

\subsubsection{User-defined Operation}
\label{sec:user-defined}
User-defined operations begin with the \quingo{operation} keyword, followed by the operation name, parameter list, return type, and finally, the operation body. The operation body is constructed in an imperative style with the basic element to be statements. With statements, it allows to freely mix quantum operations with classical operations to enable the execution of quantum operations controlled by classical program flow, such as measurement-based feedback~\cite{svore2018q}.

For simplicity, the Quingo language supports structured programming with four basic but comprehensive structures~\cite{selinger2004towards, ying2016foundations} at the core level, i.e., sequence, selection (\quingo{if-else}), loop (\quingo{while}), and recursion.
Other high-level structures, e.g., the \quingo{for} loop, can be easily constructed as syntactic sugar based on the \quingo{while} loop.
Note that classical operations in the kernel will be finally translated into classical instructions executed by the control processor, which has a fast interaction with quantum operations. Hence, conditions in the selection depending on measurement results are generated into interleaved classical and quantum instructions that are executable on nowadays hardware.

With the presented structures, the Quingo language provides more compact and readable code than some other languages also supporting the refined HQCC model. For example,
Code~\ref{code:quingo-loop} is the Quingo description which implements the same functionality using dynamic lifting as Code~\ref{code:pyquil_loop} in PyQuil. Since the Quingo language only describes tasks running on the quantum coprocessor, Code~\ref{code:quingo-loop} will be compiled into interleaved quantum and classical instructions which will be executed by the control processor to implement dynamic lifting.
No low-level data type and operations like \code{BIT} and \code{MOVE} and extra data structure like \code{Program} are required in Quingo to describe the kernel program.

\subsubsection{Operation Modifier}
Adding control qubits to operations and inverting operations can significantly improve the expressiveness of a quantum programming language. They have applications in uncomputation and many oracle-based algorithms such as phase estimation.
The Quingo language takes two keywords, \quingo{control} and \quingo{invert} to support these features.
We refer to them as \textit{operation modifiers}, and they work in a similar way as higher-order functions.
The \quingo{control} modifier takes a list of qubits and an operation as parameters and returns the controlled version of this input operation with the  qubits being the input qubit list. The \quingo{invert} modifier simply inverts the given operation.
For example, line~18 of Code~\ref{code:ipe_kernel} generates the controlled oracle using the \quingo{control} keyword. Note, the oracle in the \quingo{control} is written in such a way that it is like being called, which results in rather intuitive code.

\subsection{Timing Control}
\label{subsec:lang-timing}
\begin{figure}[bt]
    \centering
    \includegraphics[width=\textwidth]{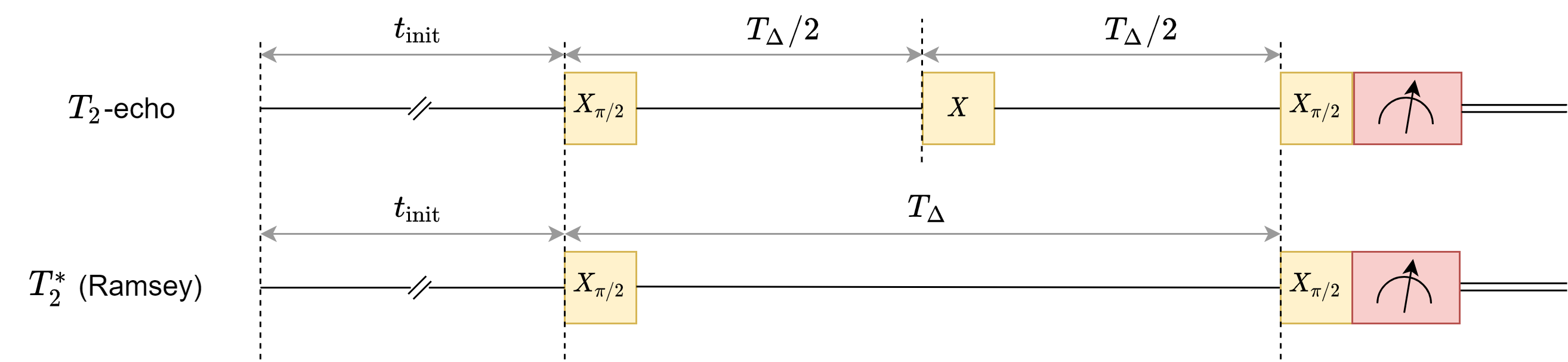}
    \caption{Timing of operations in the $\Ttwo$ experiment.}
    \label{fig:t2}
\end{figure}

As described in Section~\ref{sec:background}, experimentalists need to explicitly control the timing of operations in many quantum experiments. As experimental quantum systems have been scaled up to support more quantum algorithms, which are described at a rather low-level by the experimentalists, the boundary between quantum algorithms and experiments blurs, which calls for describing the timing of operations in algorithms.
We use the $\Ttwo$ experiment of which the operations with timing are presented in Fig.~\ref{fig:t2} to demonstrate the timing control mechanism of the Quingo language.
In the $\Ttwo$-echo experiment, three $x$-rotations ($X_{\pi/2}$, $X$, and $X_{\pi/2}$) separated by the same interval followed by a measurement are applied to the target qubit after initialization. The intervals will be set by the experimentalist according to some properties of the calibrated qubit(s), such as the dephasing time.
The $\Ttwo^*$ (Ramsey) experiment differs from the $\Ttwo$-echo experiment only by the absence of the middle $X$ gate. Code~\ref{code:t2} presents the Quingo code that describes the measurement of $\Ttwo$-echo or $\Ttwo^*$ selected by the boolean variable \quingo{echo}.

\subsubsection{Timer-based Mechanism}
Quingo introduces a timer-based scheme for controlling the timing of quantum operations. Two special types, \quingo{time} and \quingo{timer}, are added to the type system.

A \quingo{time} variable consists of a \quingo{double}-type value and a time unit such as \textit{ns} (nanosecond), which can be used to describe a timespan. There are three ways to obtain a \quingo{time}-type value: (1) a \quingo{time} literal, (2) adding or subtracting two \quingo{time} values, and (3) scaling a \quingo{time} value by an \quingo{int} or \quingo{double} value. Line~13 in Code~\ref{code:t2} shows an example of (3).

A timer, defined with the \quingo{timer} keyword (line~6 in Code~\ref{code:t2}), is a clock starting at a particular point. Quingo timers can be viewed as a special kind of external resource to the program. They advance automatically at the same pace regardless of the program's execution path. Timers can only be read or reset. Timers can be compared against \quingo{time} values to form timing constraints. Multiple timing constraints can be combined with logical operators \quingo{&&}. Timers are implicitly reset upon definition.

\begin{lstfloat}
\begin{lstlisting}[
  style=QuingoStyle,
  numbers=left,
  linewidth=0.94\columnwidth,
  xleftmargin=0.06\columnwidth,
  basicstyle=\linespread{1}\scriptsize\ttfamily,
  caption=Quingo code for the $\Ttwo$-echo and $\Ttwo^*$ (Ramsey) experiment,
  label=code:t2,
]
import operations.*
import config.json

operation t2(intervals: time[], echo: bool) : bool[] {
    bool[intervals.length] results;
    timer tmr;
    using (q : qubit) {
        for (int i = 0; i < intervals.length; i += 1) {
            init(q);
            X(q, PI/2) !{tmr};
            // Conditionally apply an X gate if this is the T2-echo experiment
            if (echo) {
                X(q, PI) @{tmr == intervals[i]/2};
            }
            // The X_pi/2 operation below gets executed at the same time point
            // no matter if the X_pi operation is executed or not.
            X(q, PI/2) @{tmr == intervals[i]} !{tmr};

            results[i] = measure(q) @{tmr == duration(X)};
        }
    }
    return results;
}
\end{lstlisting}
\end{lstfloat}

Timing of quantum operations are specified using the syntax structure:
\begin{center}
    \ttfamily
    <operation> \textcolor{blue}{@\{}<timing-constraints>\textcolor{blue}{\} !\{}<timer-list>\textcolor{blue}{\}}
\end{center}
The semantics of this structure is that the operation starts its execution at a time point at which the given timing constraints enclosed in {\ttfamily\small \textcolor{blue}{@\{...\}}} are satisfied, and at exactly the same moment, all timers enclosed in {\ttfamily\small \textcolor{blue}{!\{...\}}} get reset.  {\ttfamily\small<timing-constraint>} and {\ttfamily\small<timer-list>} are both optional in a statement.

In Code~\ref{code:t2}, the \quingo{X(q, PI/2)} operation in line~10 resets the timer \quingo{tmr}.
If the experiment measures $\Ttwo$-echo, the \quingo{X(q, PI)} operation in line~13 starts execution when the timer \quingo{tmr} reaches the time specified by \quingo{ intervals[i]/2}. Hence, an interval of \quingo{ intervals[i]/2} is inserted between the previous $X_{\pi/2}$ and this $X$ operation.
In the same way, the  operation $X_{\pi/2}$ specified by line~17 starts at the time \quingo{(interval[i])} after the first $X_{\pi/2}$.
The \quingo{duration} function is a Quingo intrinsic retrieving the duration of the given opaque operation.
With the assistance of \quingo{duration},
operations executed back-to-back can be specified.
For example, the measurement operation in line~19 starts right after the \quingo{X(q, PI/2)} operation in line~17 finishes.
The same timer can appear in both the {\ttfamily\small<timing-constraints>} and {\ttfamily\small<timer-list>} parts (e.g., \quingo{tmr} in line~18), which means this timer gets reset immediately after it reaches a time point when all listed timing constraints are satisfied.
Note that timing constraints can only be associated with quantum operations.

\subsubsection{Solving the Timing Constraint}
Through the timer-based scheme, the relative distance of different quantum operations is specified, either as a concrete value (e.g. \quingo{tmr==interval[i]/2}) or a range (e.g., \quingo{tmr>interval}). However, the underlying quantum architecture, e.g., eQASM~\cite{fu2019eqasm}, requires a determined schedule in which every operation is associated with an absolute execution time. The compiler is responsible for choosing a schedule where all timing constraints in the quantum kernel are satisfied.
Since the timing constraint can be not so strict, the same quantum operation is allowed to be applied at one of multiple timing points, and the Quingo language turns to be a non-deterministic language.

This scheduling problem can be solved by an augmented version of the list-scheduling algorithm~\cite{baker1974} commonly used by classical compilers. The \quingo{time} and \quingo{timer} variables involved in the \code{<timing-constraints>}
are modeled as read dependencies. Timers listed in the \code{<timer-list>} are modeled as write dependencies.
The timing constraints are modeled as extra latency in addition to the operation duration. By adding these dependencies to the dependency graph, the list-scheduling algorithm will determine the related order of the operations, as well as compute the execution time for each quantum operation. Since timing constraints can be non-deterministic, there could be more than one legal schedule, and heuristics can be used to make the decision.
If no feasible schedule can be found, the compiler will raise an error asking the programmer to modify the timing constraints in the program.

\subsection{Data Exchange}
\label{sec:quingo_data_ex}
Data exchange is required between the host program and Quingo kernel.
According to the six-phase quantum program life-cycle model, the host program is required to pass parameters to the Quingo kernel at the language level, and the kernel execution result should be returned to the host program.
The Quingo framework defines a protocol to enable such data exchange with nowadays available technology.
Since the quantum coprocessor can only return binary data via the shared memory to the host language, a binary format should be defined for the quantum computation result.
With the well-defined data exchange protocol between the classical host and the quantum kernel, it opens the door for replacing the classical host language or the quantum kernel language, and the compiler with other classical or quantum language or compiler by supporting the communication protocol.

\subsubsection{Host Program to Quingo Kernel}
\label{sec:read_data}

To decouple the data passing process from the compiler implementation, we propose a Quingo-source-file-based method, which contains two parts. First, the runtime system provides a set of interfaces in the host language, which can encode kernel interface parameters to a format defined by the data converter.
The programmer is responsible for preparing the kernel interface parameters into the format by calling functions in the interface, as is done in Q\#~\cite{svore2018q}.
However, if the host language supports type inference, these functions are not required to be exposed to the programmer, and the underlying implementation of the \pyth{call_quingo} function can perform the conversion automatically, which simplifies the programming in the host language. The host program may pass invalid parameters, e.g., a list of elements with different types, which is unsupported by Quingo where only homogeneous arrays are supported. In this case, errors are raised during the conversion.

The converted data is written into a generated Quingo file. This file is added to the compilation and hence can be read by the compiler.
For example, assume the host program calls kernel \quingo{iqe} with a parameter 5 in Code~\ref{code:ipe_host}. The runtime system generates file \textit{main.qu} as shown in Code~\ref{code:main}. This file contains a \quingo{main} operation that has zero parameters. Its return type is the same as the called Quingo kernel \quingo{iqe}. Then, the compiler can read this file and retrieve the kernel interface parameters to compile the quantum kernel.
\begin{lstfloat}[bt]
\begin{lstlisting}[
style=QuingoStyle,
numbers=left,
linewidth=0.94\columnwidth,
xleftmargin=0.06\columnwidth,
basicstyle=\linespread{1.0}\scriptsize\ttfamily,
escapechar=|,
label=code:main,
caption=\textit{main.qu} generated by the runtime system when Quingo kernel \code{ipe} is called.
]
operation main(): double {
   return ipe(5);
}
\end{lstlisting}
\end{lstfloat}
The advantage of this method is to simplify the design of the Quingo compiler. Its inputs are all Quingo source files, and it does not need to provide a dedicated interface for the host program (cf. \textit{Principle 3}).

\subsubsection{Quingo Kernel to the Host Program}
After the execution of a quantum program, the results are transferred back to the host by the runtime system. A shared memory space that can be accessed by both the control processor and the host is used for this data transfer. First, the Quingo kernel writes the return values to the shared memory. Next, the runtime system copies the data to the host machine. Finally, the runtime system interprets the data as the types of the host language and passes them to the host program.

The process of converting an object into a stream is normally referred to as \textit{serialization}~\cite{sumaray2012comparison}. The byte stream that the Quingo kernel writes to the shared memory can be seen as the serialization of the return data.
We propose a set of rules to define the format of the serialized data:
\begin{itemize}
    \item For data of a primitive type, write the data in a little-endian style, i.e., the least significant bits are stored in the lowest address. The \quingo{bool} values \quingo{true} and \quingo{false} are represented by 1 and 0, respectively. \quingo{double} numbers are serialized following the IEEE-745 single-precision floating-point standard.
    \item For tuple types, their elements are serialized individually, and then the results are combined.
    \item An array is serialized into an integer value that is the offset from the current address to the actual storing region. The actual storing region starts with an integer value indicating the number of the elements in the array. The serialization result of the elements is placed after the integer.
\end{itemize}
Note that the data type is not stored in the serialization result because the runtime system can fetch the data type from the Quingo kernel source program.

This data serialization format has the advantage that the content is independent of absolute memory addresses. Note that the actual array storage location is accessed with a relative address offset. This format allows the serialized byte stream to be copied to any absolute address and maintains the same semantics. This feature is essential for the HQCC model, as the host and the control processor have different memory spaces, and the serialization result needs to be transferred from the control processor to the host.

\section{Implementation \& Examples}
\label{sec:example}

\subsection{Implementation}
The runtime system embodies the Quingo framework. We have implemented\footnote{The code of the runtime system can be found at \textcolor{blue}{\url{https://github.com/quingo/runtime_system}}.} most of the runtime system as an open-source library with $\sim$1600 lines of Python code, including all modules except the pulse generator in the \textit{Runtime System} block in Fig.~\ref{fig:rtsys}.
Pulse generators and hardware drivers for real quantum devices are available in prevailing open-source quantum control environments, such as PycQED~\cite{pycqed}. They can be integrated into the runtime system using the reserved interface when the Quingo framework is connected to actual quantum setups.

An open-source prototype of the Quingo compiler\footnote{The code of the prototype compiler can be found at \textcolor{blue}{\url{https://github.com/quingo/compiler_xtext}}.} has been implemented using the Xtext~\cite{eysholdt2010xtext} framework with $\sim$200 lines of Xtext code, $\sim$700 lines of Xsemantics~\cite{Bettini2016Implementing} code, and $\sim$6000 lines of Xtend code.
The prototype compiler comprises a front end and a code generator.
The front end is automatically generated from a description of the Quingo syntax in Xtext and the Quingo semantics in Xsemantics~\cite{Bettini2016Implementing}.
The output of the front end is an abstract syntax tree of the Quingo program, which forms the input to the code generator.
While traversing the nodes of the abstract syntax tree, the code generator partially executes the quantum kernel using techniques such as constant propagation and dead code elimination. This step can simplify the abstract syntax tree of the quantum kernel by removing nodes whose values are known at static time.
For nodes corresponding to quantum operations or classical operations whose operands are unknown at compile time (e.g., the value comes from a measurement of a qubit), the code generator emits corresponding quantum or classical eQASM instructions, which will be eventually executed by the control processor.

eQASM programs can be simulated by the quantum control architecture simulator CACTUS~\cite{fu2020cactus} connected to a quantum state simulator, like QuantumSim~\cite{brien2017density}. Since CACTUS is a cycle-accurate simulator, of which the simulation speed is relatively low, we developed a functional simulator, PyCACTUS~\footnote{The code of PyCACTUS can be found at \textcolor{blue}{\url{https://github.com/gtaifu/PyCACTUS}}.} for eQASM. PyCACTUS is implemented using $\sim$2000 lines of Python code.

We have also implemented tens of examples of Quingo applications with the host language being Python including some algorithms and quantum experiments~\footnote{ The Quingo examples can be found at \textcolor{blue}{\url{https://github.com/quingo/quingo_examples}}.}.

Due to a lack of a linear intermediate representation like LLVM~\cite{lattner2004llvm} in the prototype compiler, it is difficult to perform global analysis (including timing analysis of quantum operations) or transformation on the program. Hence, we postpone solving the timing constraint of Quingo programs in a new compiler based on the multi-level intermediate representation (MLIR)~\cite{lattner2020mlir}, which is currently under development. Adding control qubits to or inverting a quantum operation would significantly improve the expressiveness of the Quingo language. This functionality has not been implemented due to limited manpower and will be implemented once we reach a stable enough compiler that can process the quantum-classical interaction with optimization.

\subsection{Example}

\begin{lstfloat}[bt]
\begin{lstlisting}[
style=QuingoStyle,
numbers=left,
linewidth=0.94\columnwidth,
xleftmargin=0.06\columnwidth,
basicstyle=\linespread{1.0}\scriptsize\ttfamily,
escapechar=|,
label=code:rng-kernel,
caption=Quingo kernel implementing the accumulator and random number selector.
]
// kernel.qu
import config.json.*
import operations.*

operation sum_random(arr: int[], r: bool): (int, int) {
    int sum = 0, i = 0, random = 0;
    while (i < arr.length) {
        sum += arr[i];
        i += 1;
    }

    if (r) {
        using(q: qubit) {
            init(q);  H(q);
            if (measure(q)) { random = arr[0]; }
            else { random = arr[1]; }
        }
    }
    return (sum, random);
}
\end{lstlisting}

\begin{lstlisting}[
style=mypython,
numbers=left,
linewidth=0.94\columnwidth,
xleftmargin=0.06\columnwidth,
basicstyle=\linespread{1.0}\scriptsize\ttfamily,
label=code:rng-host,
caption=Python host calling the quantum accumulator and random number selector.
]
# host.py
from qgrtsys import if_quingo
from pathlib import Path

kernel_file = Path(__file__).parent / "kernel.qu"
if not if_quingo.call_quingo(kernel_file, 'sum_random', [2, 6, 8], False):
    raise Error("failed to call the quantum kernel.")

res = if_quingo.read_result()
print("result of add example is:", res)
\end{lstlisting}

\begin{lstlisting}[
style=QuingoStyle,
numbers=left,
linewidth=0.94\columnwidth,
xleftmargin=0.06\columnwidth,
basicstyle=\linespread{1.0}\scriptsize\ttfamily,
escapechar=|,
label=code:rng-main,
caption=\code{main} operation generated by the runtime system.
]
operation main(): (int, int) {
    int[] var0_arr = {2, 6, 8};
    bool var1_bool = false;
    return sum_random(var0_arr,var1_bool);
}
\end{lstlisting}

\begin{lstlisting}[
  style=eQASMStyle,
  numbers=left,
  linewidth=0.94\columnwidth,
  xleftmargin=0.06\columnwidth,
  basicstyle=\linespread{1}\scriptsize\ttfamily,
  caption=Generated eQASM code when the condition is false.,
  label=code:random_false,
]
XOR r0, r0, r0
ADDI r1, r0, 1
LDI r2, 0x20000
LDUI r2, r2, 0x1
SW r0, 0x10000(r0)
FCVT.S.W f0, r0
# start of sum_random
# end of sum_random
ADDI r6, r0, 0      # a 'stack' pointer to the shared memory, used for exporting
ADDI r7, r0, 0      # a 'heap' pointer to the shared memory, used for exporting
# start exporting: (int,int)
ADDI r8, r0, 0      # load sharedAddr to the base register: r8
ADDI r9, r0, 16     # exporting: int (sum = 16)
SW r9, 0x0(r8)
ADDI r10, r0, 0     # exporting: int (random = 0)
SW r10, 0x4(r8)
STOP
\end{lstlisting}

\end{lstfloat}

We take an example to illustrate how components in the Quingo framework work together seamlessly. The kernel and host program of the example are shown in Code~\ref{code:rng-kernel} and Code~\ref{code:rng-host}, respectively. The kernel operation \quingo{sum_random} takes a list of at least two integers (\quingo{arr}) and a boolean value (\quingo{r}) as parameters. It returns two values. The first one is the sum of integers in the list \quingo{arr}.
If \code{r} is \quingo{false}, the second one is 0.
Otherwise, the second one is randomly chosen from \code{arr[0]} or \code{arr[1]}, with the random flag generated from measuring an equally superposed qubit (\lines{13}{15}).

After the host program calls the kernel with parameters (line~6 of Code~\ref{code:rng-host}), the runtime system converts the given parameters into the corresponding data types in Quingo, and generate a \quingo{main} operation which actually calls \quingo{sum_random} (Code~\ref{code:rng-main}). The phase manager then triggers the compiler to compile related Quingo files and configuration files. During compilation, the compiler performs optimization over the quantum kernel, and generates corresponding eQASM file (Code~\ref{code:random_false}) for execution.
Since \code{r} is \quingo{false}, the \quingo{if} body can be directly eliminated via partial execution at compile time.
Together with constant propagation, other classical operations in the kernel can also be eliminated.
As a result, there are neither classical or quantum instructions generated from \quingo{sum_random}.
Note, when \code{r} is true, corresponding quantum and classical instructions will be generated to perform dynamic selection (see Code~\ref{code:random_true}).

After the quantum coprocessor finishes computation and gets the result, it should serialize and store the result into the shared memory with a starting address of 0, which task is fulfilled by the last instructions of the eQASM program (\lines{12}{16} of Code~\ref{code:random_false}).
These special instructions are generated by the compiler when compiling the return statement of the main operation.
Later, when the host language tries to retrieve the kernel computation result using \code{if\_quingo.read\_result()} (line~9 of Code~\ref{code:rng-host}), the runtime system retrieves the leading bytes in the shared memory (result block), decodes them into the data types of the host language, and return them to the host program.
Since then, the host program can process the kernel computation result as required, such as print it (line~10 of Code~\ref{code:rng-host}).

\section{Discussion}
\label{sec:rr}

\subsection{Quantum Experiment Support of Quingo \textit{v.s.} Other QPLs}
The core goal of the Quingo language is to support quantum experiments and assist NISQ algorithms, which differentiates Quingo from other QPLs or programming framework.
To this end, the Quingo language introduces mechanisms to support interacting with low-level details, including the usage of opaque operations and the timing control of operations. As a result, the Quingo language becomes a QPL that is not so high level compared with many other QPLs.

Timing control plays a key role in quantum experiments. Nevertheless, almost all existing QPLs actively neglect the requirement on timing control since it is low-level hardware detail.
An exception is OpenQL, which supports controlling the timing of quantum operations using the wait statements.
However, wait statements can be cumbersome when used to describe the timing of multiple qubits if program flow controls such as loops are required. Though being straightforward for a small number of qubits, the complete semantics of wait statements is difficult to comprehend, which may result in unexpected compilation results.
The Quingo language proposed a timer-based timing control scheme at the language level, which is more flexible than the wait-statement-based timing control.
The underlying model is the timed automata proposed by~\cite{alur1994theory}, of which the semantics related to timing control is well defined, easy to understand, even with the presence of classical constructs. As a result, the program with complex timing control could also be easier to write.

The semantics of quantum operations in a quantum program is assumed to be well-defined in most QPLs, although their implementation might be opaque.
The Quingo language supports the usage of opaque operations without well-defined quantum semantics and treats them as pulse(s) applying on one or multiple qubits with a certain duration.
A dedicated configuration system is coupled with the Quingo language to bind opaque operations to concrete quantum semantics or particular pulses, and opaque operations will be treated as black boxes during compilation if no quantum semantics is provided.

\subsection{Embedded \textit{v.s.} External DSL for HQCC}
\label{sec:edsl}
While most quantum programming languages are implemented as eDSLs, such as Quipper, ProjectQ, Qiskit, PyQuil, Cirq, and OpenQL, the Quingo framework advocates external DSLs for quantum computing.
The rationale is the intrinsic disadvantage of using eDSLs to describe real-time quantum-classical interaction, which deeply roots in the life cycle of an eDSL-based quantum program.

\begin{figure}[bt]
    \centering
    \includegraphics[width=\textwidth]{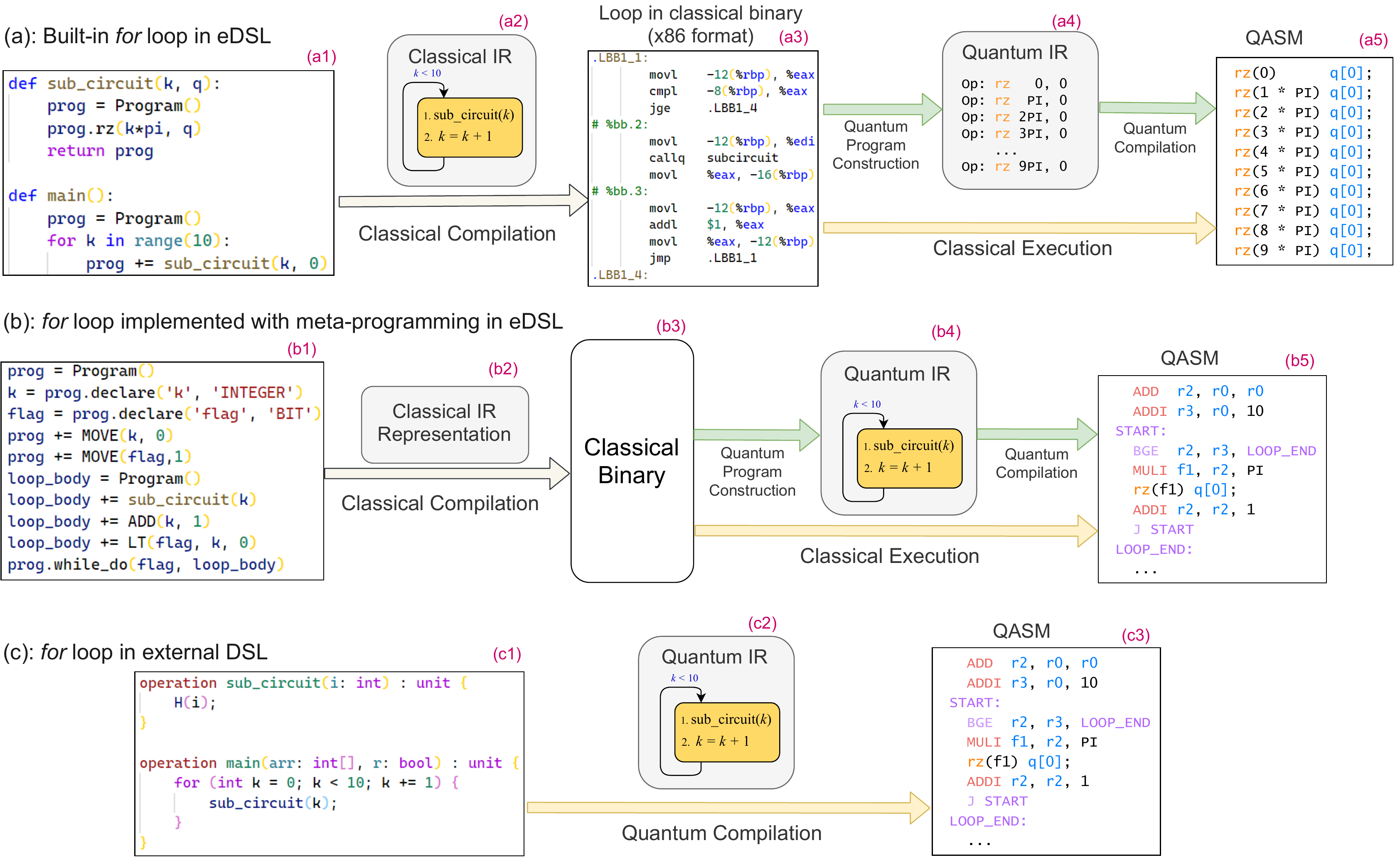}
    \caption{Compilation process of quantum programs with various loop implementations in an eDSL or external DSL. (a) Loop implemented directly using built-in flow control structures in the host language of an eDSL; (b) Loop implemented using meta-programming techniques in an eDSL; (c) Loop implemented using flow control structures in an external DSL.}
    \label{fig:edsl-for}
\end{figure}

By analyzing existing eDSLs for quantum computing, we observe that eDSLs usually provide a library in the host language with an interface of several methods including:
\begin{enumerate}[label=(\roman*)]
    \item instantiating objects representing quantum circuits or programs,
    \item adding operations or sub-circuits to construct the quantum program, and
    \item compiling and executing the quantum program.
\end{enumerate}
After being written with these methods, an eDSL-based quantum program is first compiled into a classical binary by a classical compiler and then executed by the classical computer\footnote{For the sake of simplicity, we blur the difference between compiled languages and interpreted languages here and in Fig.~\ref{fig:edsl-for}. The difference in these two kinds of languages does not affect the conclusion of this subsection. }.
The classical execution stage starts with some classical pre-processing followed by the construction of the kernel program.
Thereafter, the quantum compiler is triggered to optimize the kernel program.
A quantum binary forms the final output of both the quantum compilation stage and classical execution stage, which is sent to the quantum coprocessor for execution.
The life cycle of an eDSL-based quantum program is shown in Fig.~\ref{fig:edsl-for}(a) (the classical pre-processing is not present in the classical execution stage for simplicity).

However, built-in classical operations and control flow structures provided by the host language, such as the addition operation and loops, \textit{cannot} be translated into classical instructions in the quantum binary.
Take the loop as shown in Fig.~\ref{fig:edsl-for}(a) as an example.
The loop structure in the host program ({a1}) will be parsed into a classical IR format ({a2}) by the classical compiler and translated into classical instructions ({a3}) including \code{cmpl}, \code{jge}, \code{jmp}, and so on.
These classical instructions are executed by the classical computer, resulting in multiple occurrences of the sub-circuit with corresponding parameters in the quantum IR ({a4}), and hence multiple \code{rz} gates in the quantum binary ({a5}).
That no classical instructions can be directly generated from the host language primitives forms an obstacle in describing real-time quantum-classical interaction when using eDSLs for quantum computing.

The quantum-IR-format classical operations or control flow structures are the keys to generate classical instructions in the quantum binary.
eDSLs like PyQuil introduces meta-programming techniques such as the \code{BIT} data type and \code{while\_do} function to build classical operations or control flow structures in the quantum IR.
As shown in Fig.~\ref{fig:edsl-for}(b), the eDSL source code utilizing meta-programming ({b1}) is parsed into classical IR ({b2}) and translated into classical binary ({b3}).
Classical operations and control flow structure introduced by meta-programming will only be constructed using dedicated data structure after the classical binary is executed.
Hence, these constructs can go through the classical binary layer and appear in the quantum IR (b4), which are finally translated into classical instructions in the quantum binary ({b5}).
However, this inevitably increases the code complexity and reduces its readability.
Such inefficiency is clearly demonstrated by the PyQuil implementation (Code.~\ref{code:pyquil-ipe}) of the same IPE algorithm as the Quingo description (Code~\ref{code:ipe_kernel}), where the former costs 44 lines of code while the latter only 24 lines.
Also, describing structured programs with meta-programming in eDSLs is much less readable than external DSLs. A reader can hardly recognize the loop structure with its various components within \lines{26}{40} of Code~\ref{code:pyquil-ipe} at first glance.

To enable fast quantum-classical interaction and avoid complex source code, the Quingo framework advocates using a classical language for the host program and an external DSL merely for the quantum kernel.
As NISQ qubits have very short coherence time, it is unlikely to interpret and execute a quantum program on NISQ hardware.
Hence, it is natural for QPLs to be compiled languages other than interpreted languages.
As shown in Fig.~\ref{fig:edsl-for}(c), the quantum compiler can parse the quantum kernel described in an external DSL into the quantum IR directly.
In this way, classical operations and control flow structures can be kept in the quantum IR and finally translated into classical instructions in the quantum binary.
Both the Q\# language and Quingo language adopt this design.

Another advantage of external DSL is that it is not restricted by the syntax of an existing language. For example, the \quingo{!} operator in the Quingo language is used to denote resetting timers in the following braces. This freedom is hard, if not impossible, to achieve when using an eDSL.

\subsection{HQCC Support of Quingo \textit{v.s.} Q\#}

The Q\# language is also an external DSL and adopts a model similar to the refined HQCC model including three levels, i.e., the off-line classical computation, quantum computation, and real-time classical computation~\cite{svore2018q}.
Since Q\# aims to target large-scale quantum computation for future hardware, there is no assumption about any hardware details, such as the latency difference between two kinds of quantum-classical interaction, and the capability of the control processor.
Hence, programmers are not explicitly guided to put heavy classical tasks without real-time interaction with quantum operation in the classical host when programming with Q\#.
Take as an example the function \code{EstimatePhase} which estimates the phase of an oracle iteratively~\cite{qs2020ipe}.
The only line of code in this function for quantum computation is calling the function \code{ApplyIterativePhaseEstimationStep}, which is sandwiched between classical tasks including integration and array manipulation which does not require real-time interaction with quantum operations.
In contrast, the control processor in the refined HQCC model would have limited classical power.
Correspondingly, the Quingo language is designed with moderate classical processing capability.
Hence, it can form an implicit guide for the programmer to put classical tasks not requiring fast interaction with qubits to the classical computer.

\subsection{Program Life Cycle of Quingo \textit{v.s.} Quipper}
Based on the restricted HQCC model, the Quipper program life cycle is divided into three phases, including the compile time, circuit generation time, and circuit execution time (see section 4.3.1 in~\cite{green2013quipper}).
Classical operations in Quipper can be only performed by the classical computer at circuit generation time, which cannot have fast interactions with quantum operations which are executed at circuit execution time.
To support dynamic lifting, it needs to assume that there is long-term storage that can preserve qubits during the alternation between the circuit generation time and circuit execution time.
This life cycle is not suitable for programs based on the refined HQCC model where fast interaction between classical and quantum operations is available.
By utilizing the fast interaction between the control processor and qubits, the six-phase quantum program life-cycle model can naturally support dynamic lifting without assuming the long-term storage for qubits.
As a result, the six-phase quantum program life-cycle model can naturally support dynamic lifting with nowadays available hardware.

\subsection{Optimization via Semi-Static Compilation}
Semi-static compilation for quantum kernels can further exploit the potential of HQCC architectures to support HQCC.
Semi-static compilation utilizes classical computation power to optimize the quantum kernel, which can help improve the fidelity of quantum algorithms.

Semi-static compilation offloads the classical instructions that should be executed by the less powerful control processor to the more powerful classical host. For example, the compiler running on a classical computer performs aggressive optimization over the  \quingo{sum_random} operation, resulting in zero instructions that should be executed on the quantum coprocessor for this operation, as shown by \lines{7}{8} in Code~\ref{code:random_false}.
Fewer operations executed by the control processor can lead to two advantages in the NISQ era.
First, with a possibly shorter quantum execution time, which is highly dependent on the control processor execution time, the fidelity of the quantum program could be improved.
Second, the control processor has limited processing power and memory space due to resource constraints. For example, most nowadays control processors or control devices serving as a control processor are implemented with dedicated soft cores on FPGA~\cite{fu2017experimental, fu2018microarchitecture, fu2019eqasm, ryan2017hardware, bourdeauducq2018artiq, keysightm3202a, zi2030pqsc}.
The semi-static compilation can offload some classical computation to the classical host via partial execution, which enables the programmer to describe the quantum kernel with more classical computation power than what the control processor can offer.

The idea of performing aggressive optimization over the quantum program based on static analysis and partial execution was first proposed by~\cite{javadiabhari2015scaffcc} with an implementation in ScaffCC based on LLVM~\cite{lattner2004llvm}.
Since its design bears no control processor in mind, loop unrolling and procedure cloning are extensively used during the compilation. The compiler cannot generate reliable code describing quantum-classical interaction with timing constraints satisfied to support flow control in real-time, and the size of the generated quantum code can be big.

\section{Conclusion \& Future Work}
\label{sec:conclusion}

This paper summarizes three kinds of execution models that can depict quantum-classical interaction in most quantum programming languages.
By analyzing the difference among these three models, we found the refined HQCC model can best suit NISQ computing system implementation.
The refined HQCC model clarifies the difference between two kinds of quantum-classical interaction, i.e., the slow interaction between the classical host and quantum coprocessor, and the fast interaction between classical operations and quantum operations in the quantum coprocessor. To integrate and manage quantum and classical computing resources, \textbf{required is a framework more than a language for HQCC}.

This paper proposes the Quingo framework at a system level based on the refined HQCC model.
The Quingo framework enables the programmer to code HQCC applications in a neat programming model.
With higher confidence, quantum programs described in this model can be mapped to a real quantum computing system for execution with timing constraints satisfied.
By introducing a novel, six-phase quantum program life-cycle model based on the refined HQCC model,  optimization space of the quantum kernel is reserved for semi-static compilation, which could lay a foundation for co-optimization of mixed quantum and classical computation.

With flexible timing control at the language level and a mechanism for primitive operation definitions, the Quingo language can be used to describe a wide range of quantum experiments. Compared to other quantum programming languages with many high-level features, the Quingo language is a relatively low-level QPL.
The Quingo language is expected to bridge the gap between various quantum software and quantum experiments, which implies a closer description of quantum algorithms to real quantum machines.

The next steps for Quingo include:
introducing more high-level features to Quingo to ease the programming of complex quantum algorithms, such as controlled quantum gate generation and automatic uncomputation; developing a more powerful compiler for the Quingo language based on some compilation framework, such as LLVM~\cite{lattner2004llvm} or MLIR~\cite{lattner2020mlir}; integrating the Quingo framework with quantum control environments, such as PycQED~\cite{pycqed}; and demonstrating its application in quantum experiments with real hardware.

Hopefully, the Quingo framework could guide the design of an HQCC system enabling seamless collaboration between quantum and classical software and hardware in the future.

\begin{acks}
    We would like to thank Center for Quantum Computing, Peng Cheng Laboratory for their generous support.
    We thank Yizhi~Wang and Yujie~Liu for discussing NISQ algorithms, \textcolor{blue}{\url{https://git.pcl.ac.cn}} and Y. Yu for hosting Quingo-related code with help.

    This work is supported by National Natural Science Foundation of China under Grant No. 61902410 (X.F.), Grant No. 61632021 (J.W.), Grant No. 62072176 (Y.D.), and Grant No. 61832015 (Y.D.), the National Key R\&D Program of China under Grant No. 2018YFA0306704 (Y.F.) and the Australian Research Council under Grant No. DP180100691 (Y.F.).
\end{acks}

\bibliographystyle{ieeetr}
\bibliography{ms.bib}
\section{Example Quingo Code}

\begin{lstfloat}
\begin{lstlisting}[
  style=eQASMStyle,
  numbers=left,
  linewidth=0.94\columnwidth,
  xleftmargin=0.06\columnwidth,
  basicstyle=\linespread{1}\scriptsize\ttfamily,
  caption=Generated eQASM code when the condition is true.,
  label=code:random_true,
]
XOR r0, r0, r0
ADDI r1, r0, 1
LDI r2, 0x20000
LDUI r2, r2, 0x1
SW r0, 0x10000(r0)
FCVT.S.W f0, r0
# start of sum_random
SMIS s0, {0}
MeasZ s0            # start of init
FMR r3, q0
ADD r4, r3, r0
BNE r4, r1, if_0_end
rx180 s0
if_0_end:           # end of init
H s0                # H followed by msmt
MeasZ s0
FMR r5, q0          # r5 gets the random msmt result
ADDI r6, r0, 3
SW r6, 0x10022(r0)
BNE r5, r1, if_1_end
ADDI r7, r0, 5
SW r7, 0x10022(r0)
if_1_end:
LW r8, 0x10022(r0)
ADD r9, r8, r0
# end of sum_random
ADDI r10, r0, 0     # a 'stack' pointer to the shared memory, used for exporting
ADDI r11, r0, 0     # a 'heap' pointer to the shared memory, used for exporting
# start exporting: (int,int)
ADDI r12, r0, 0     # load sharedAddr to the base register: r12
ADDI r13, r0, 8     # exporting: int (c = 8)
SW r13, 0x0(r12)
SW r9, 0x4(r12)     # exporting: int (random)
STOP
\end{lstlisting}
\end{lstfloat}

\begin{lstfloat}

\begin{lstlisting}[
  style=mypython,
  numbers=left,
  linewidth=0.94\columnwidth,
  xleftmargin=0.06\columnwidth,
  basicstyle=\linespread{1}\scriptsize\ttfamily,
  caption=PyQuil implementation of IPE.,
  label=code:pyquil-ipe,
]
def pyquil_ipe(prog: Program, controlled_oracle, m):
    eigenstate = 1
    ancilla = 0

    power = prog.declare('power', 'REAL')
    theta = prog.declare('theta', 'REAL')
    prog += MOVE(theta, 0.0)
    prog += MOVE(power, float(2 ** (m - 1)))

    ro_es = prog.declare('ro_es', 'BIT')  # prepare the eigenstate |1>
    prog += MEASURE(eigenstate, ro_es)
    prog += NOT(ro_es)
    prog.if_then(ro_es, Program(X(eigenstate)))

    ro_ancilla = prog.declare('ro_ancilla', 'BIT')
    prog += MEASURE(ancilla, ro_ancilla)

    k = prog.declare('k', 'INTEGER')  # start of for Loop
    prog += MOVE(k, m)
    flag = prog.declare('flag', 'BIT')
    prog += GT(flag, k, 0)

    loop_body = Program()
    prog.if_then(ro_ancilla, Program(X(ancilla)))  # reset ancilla to |0>

    loop_body += H(ancilla)
    loop_body += controlled_oracle(ancilla, eigenstate, power)
    loop_body += PHASE(theta, ancilla)  # Z(theta_k)
    loop_body += H(ancilla)
    loop_body += MEASURE(ancilla, ro_ancilla)

    loop_body += DIV(theta, 2)
    if_branch = Program(ADD(theta, -0.5 * pi))
    loop_body.if_then(ro_ancilla, if_branch)

    loop_body += DIV(power, 2)

    loop_body += SUB(k, 1)
    loop_body += GE(flag, k, 0)
    prog.while_do(flag, loop_body)  # end of for loop

    ro = prog.declare('ro', 'REAL')  # try to return the estimated phase
    prog += MUL(theta, -1)
    prog += MOVE(ro, theta)
\end{lstlisting}

\end{lstfloat}

\end{document}